\documentclass[trackchanges,twocolumn]{aastex701}

\newcommand{\citeg}[1]{\citep[e.g.,][]{#1}}
\newcommand{\citesee}[1]{\citep[see,][]{#1}}

\newcommand{\kms}{\ensuremath{\mathrm{km\,s^{-1}}}}
\newcommand{\msun}{\ensuremath{M_\odot}}

\usepackage{hyperref} 
\usepackage{hypcap}

\newcommand{\Kharthirteen}{\cite{2013A&A...558A..53K}}
\newcommand{\GStwentyone}{\cite{2021A&A...656A..51G}}
\newcommand{\HRtwentythree}{\cite{2023A&A...673A.114H}}
\newcommand{\Mintwentyfive}{\citep{2025A&A...702A..79M}}
\newcommand{\Weidtwenty}{\cite{2020A&A...640A..10W}}
\newcommand{\Carninteen}{\cite{2019A&A...623A..80C}}

\usepackage{etoolbox}

\begin{document}

\title{Planetary Nebula Central Stars as Tracers of Planetary Nebula–Star Cluster Associations in the Galaxy}

\author[orcid=0000-0002-8634-4204]{Vasiliki Fragkou}
\affiliation{Observatório do Valongo, Universidade Federal do Rio de Janeiro \\
Ladeira do Pedro Antônio-43 \\
Rio de Janeiro, 20080-090, Brazil}
\email{vfragkou@ov.ufrj.br}  

\correspondingauthor{Vasiliki Fragkou}
\email{vfragkou@ov.ufrj.br}

\author[orcid=0000-0003-2127-2841]{Luis Lomel\'i-N\'u\~nez} 
\affiliation{Observatório do Valongo, Universidade Federal do Rio de Janeiro \\
Ladeira do Pedro Antônio-43 \\
Rio de Janeiro, 20080-090, Brazil}
\email{luislomeli@ov.ufrj.br}

\author[orcid=0000-0002-7064-099X]{Dante Minniti} 
\affiliation{Instituto de Astrof\'isica, Depto. de. F\'isica y Astronom\'ia, Facultad de Ciencias Exactas, Universidad Andr\'es Bello, Av. Fern\'andez Concha 700, Las Condes, Santiago, Chile}
\affiliation{Vatican Observatory, V00120 Vatican City State, Italy}
\email{vvvdante@gmail.com}

\author[orcid=0000-0002-0620-136X]{Arianna Cortesi} 
\affiliation{Observatório do Valongo, Universidade Federal do Rio de Janeiro \\
Ladeira do Pedro Antônio-43 \\
Rio de Janeiro, 20080-090, Brazil}
\affiliation{Instituto de Física, Universidade Federal do Rio de Janeiro, 21941-972, Rio de Janeiro, RJ, Brazil}
\email{aricorte@ov.ufrj.br}

\author[orcid=0000-0001-9388-7146]{Denise R. Gonçalves} 
\affiliation{Observatório do Valongo, Universidade Federal do Rio de Janeiro \\
Ladeira do Pedro Antônio-43 \\
Rio de Janeiro, 20080-090, Brazil}
\email{denise@ov.ufrj.br}

\begin{abstract}

Planetary nebulae (PNe) that are physically associated with star clusters are rare but crucial for stellar evolution studies because they allow their initial and final stellar properties to be determined independently, contributing to constraints on the initial-to-final-mass relation (IFMR), which traces stellar mass loss. Although the identification of such pairs is essential, most reported cases have been identified incidentally rather than through systematic searches. Here, we attempt to uncover new such associations in our Galaxy employing literature PN central stars (CSs) and star cluster catalogues. Exploring relevant angular positions on the sky and kinematic Gaia DR3 data, we identify 38 candidate CS-cluster pairs. The components of eight of these pairs present consistent interstellar extinctions, and their CS properties are compatible with stellar evolution models. Two of these candidate pairs are already reported in the literature as true associations. We evaluate separately the remaining six pairs using literature PN observational data, assessing the likelihood of each pair being a true association. The initial and final masses of our best candidates are mostly consistent with the latest IFMR predictions and confirmed cluster-PNe points. An additional 46 CS-cluster pairs have components with compatible angular positions but lack kinematic CS data and are thus presented separately. More data are needed for a precise estimation of the physical parameters of most of our candidates and to confirm these associations, which will be addressed in future studies.

\end{abstract}


\section{Introduction} \label{sec:intro}

Planetary nebulae (PNe) are a short-lived phase of the final stages of evolution of low-to-intermediate mass (1-8 \msun) stars. They consist of the material (gas and dust) ejected during the asymptotic giant branch (AGB) phase and the central star (CS), which is the hot stellar remnant, rapidly evolving to a white dwarf (WD) and capable of ionizing the surrounding material \citep{1982ApJ...258..280K}. PNe are bright emission-line sources and can be detected at larger distances than their stellar progenitors through their strong optical lines, while also emitting in X-rays, ultraviolet (UV), infrared (IR), and radio wavelengths. Due to the prevalence of the PN phase and its role, their study is essential for our understanding of the physical processes behind the late-stage stellar evolution of low-to-intermediate mass stars and the chemical enrichment of the interstellar medium (ISM). 

CSs are the PNe cores, usually (but not always) located close to the nebular geometric centre, with effective temperatures higher than 25 kK, which is the temperature typically required for the complete ionization of hydrogen in the nebula \citep{2000oepn.book.....K}. They are strong UV emitters and their effective temperatures increase until they reach the WD cooling track, while their luminosities remain constant before starting to rapidly decrease when entering the WD cooling track \citep{2016A&A...588A..25M}. Their direct study and observational and physical properties can provide important information regarding the fate of low-to-intermediate mass stars. Although CSs are expected to present bluer colours compared to their surrounding stars, their detection is often a challenging task, with many misidentifications in the literature, especially in cases where the CSs are in crowded fields, too faint, or far from the nebular geometric centre \citep{2021A&A...656A..51G, 2022Galax..10...32P, 2023MNRAS.520..773R}. In recent years, there have been continuous efforts by multiple authors \citep{2020A&A...640A..10W, 2020A&A...638A.103C, 2021A&A...656A..51G, 2022Galax..10...32P} to compile clean CS catalogues, removing such misidentifications (see Section \ref{sec:cat}). 

Star clusters are stellar populations formed from the same molecular cloud \citesee{2003ARA&A..41...57L}. Cluster stars are gravitationally bound, and since they have the same origin, they generally share similar kinematic and physical properties, such as distances, ages, and metallicities. Since cluster stars have typically the same age, it is possible to estimate their current turn-off mass, namely the mass they have upon leaving the main-sequence (MS). In our Galaxy, star clusters are traditionally categorized as old, dense globular clusters (GCs) and looser open clusters (OCs) that can be significantly younger. In the Milky Way, only slightly more than 200 GCs have been identified \citep{1996AJ....112.1487H,2024A&A...687A.201B,2024A&A...687A.214G}, with this relatively small number explained by the fact that GCs have been mostly formed in the early Universe \citeg{2013ApJ...775..134V} and only the most massive of them have survived until today \citeg{2019ARA&A..57..227K}. Although younger and continuously forming OCs are expected to be much more numerous than GCs in our Galaxy, it is estimated that only 4\% of their true number is currently uncovered \citep{2024A&A...686A..42H}, with their bulk residing in the Galaxy's thin disk \citeg{2013A&A...558A..53K}. Kinematic data obtained with the Gaia satellite \citep{2022A&A...667A.148G} provide an excellent opportunity for both the identification of new star clusters in our Galaxy and the compilation of a clean sample, and many recent studies \citep[e.g.,][and references therein]{2024A&A...686A..42H} have undertaken this task (see Section \ref{sec:cat}). Although some known OCs not uncovered with Gaia could be misidentifications, some of these, especially those detected in the IR wavelengths, may be hidden from Gaia due to interstellar extinction in the optical wavelengths \citep{2020A&A...633A..99C}. For the same reason, Gaia data fail to provide a complete sample of OCs, since many of these are inaccessible due to extinction \citep{2023ApJS..264....8H}, but it is currently the best available dataset for OC detections. 

PNe that are physically associated with star clusters provide a unique opportunity to link the observable PNe properties to the properties of their progenitors, determined from cluster studies and theoretical isochrone fitting. Most importantly, they contribute to the estimation of the initial-to-final mass relation (IFMR; \citealp[e.g.,][]{2018ApJ...866...21C}), usually constrained from cluster WDs, that is essential for stellar evolution studies and the enrichment of the ISM \citesee{2019PhDT........68F}. Unfortunately, such secure associations are very rare, with only four PNe commonly reported in the literature as confirmed to be physically associated with GCs \citep{1928PASP...40..342P, 1989ApJ...338..862G, 1997AJ....114.2611J} and five with OCs \citep{2011MNRAS.413.1835P, 2019NatAs...3..851F, 2022ApJ...935L..35F, 2025A&A...696A.146F, 2026ApJ...996...90F} in our Galaxy, with some PNe in OCs presenting extreme properties, such as large kinematic ages and relatively high-mass progenitors for a PN. Furthermore, a PN has been confirmed to be associated with a massive young cluster in the Large Magellanic Cloud \citep{2025PASP..137k4202B}. Recently, both the true nature of the PN GJJC~1, which is reported as a member of NGC~6656 (see Section \ref{sub:ind}; \citealp{2017ApJ...836...93J}), and the association of PN JaFu~1 with Palomar~6 \citep{2024AJ....168..160B} have been debated. Both these clusters are classified as GCs. In addition, multiple authors have identified a number of candidate cluster-PNe, both in our Galaxy and beyond \citep{2002ApJ...575L..59M,2008A&A...477L..17L,2013ApJ...769...10J,2014A&A...561A.119M,2019ApJ...884..115D,2019PhDT........68F,2019ApJ...884L..15M, 2021A&A...650L..11M}.

In this work, we use literature CS and cluster catalogues, along with their kinematic and physical properties, to perform a systematic search for new candidate cluster-PN associations. In Section \ref{sec:cat} we describe the source catalogues, while Section \ref{sec:res} presents our method and results. In Section \ref{sec:ext} we further refine our sample based on extinction and evolutionary constraints, and in Section \ref{sec:dis} we discuss our findings. Finally, in Section \ref{sec:conc} we summarize our conclusions.

\section{Source Catalogues} \label{sec:cat}

We employ the CS catalogue of \cite{2021A&A...656A..51G} that is the most extensive to date (3257 entries in total), containing CS candidates uncovered using Gaia EDR3 data, their dereddened BP-RP colour (indicating temperature) and stellar distances from the nebular geometric centres. Their sample is divided into groups A, B and C in order of reliability (with group A candidates being the most reliable identifications), with 2035 CS candidates in groups A and B and 850 in group C. For the purposes of our study, we assigned a CS\_ID to each candidate, which is used as an identifier throughout the following analysis. \cite{2025MNRAS.543.3035C} examined the reliability of a sample of CSs in the catalogue of \cite{2021A&A...656A..51G}, based on their luminosities, spectral energy distributions (SEDs), and other criteria being compatible with hot stars and stellar evolution models. They found that fractions of 25\% of the \cite{2021A&A...656A..51G} combined groups A and B, and 50\% of their group C, are estimated to be misidentifications, especially in the cases where the true CS is too faint to be reachable with Gaia \citeg{2026ApJ...996...90F}.

For star clusters, we adopted the catalogues of \cite{2023A&A...673A.114H} (as updated by \citealp{2024A&A...686A..42H}), \cite{2013A&A...558A..53K}, \cite{2023ApJS..264....8H}, \cite{2023ApJS..265...12Q}, \cite{1996AJ....112.1487H} \footnote{last updated 6 August 2015}, and \cite{2021MNRAS.505.5978V}. In the calculations described in Sections \ref{sec:res} and \ref{sec:ext}, for clusters with independent parameter estimates in multiple catalogues (e.g. parallax, age, extinction, metallicity), we adopted the parameters according to the catalogue sequence listed at the beginning of this paragraph. In the case of clusters that have distance but not parallax estimates (like those in the \citealp{2013A&A...558A..53K} catalogue), we convert the distances $d$ to parallaxes as \(\varpi~(\mathrm{mas}) = \frac{1}{d~(\mathrm{kpc})} \). When the source catalogue includes the cluster reddening instead of the visual extinction, we convert to extinction as $A_V=3.1E(B-V)$ \citep{1989ApJ...345..245C}.  

\cite{2023A&A...673A.114H} conducted a blind search using Gaia DR3 data to detect 7167 star clusters in our Galaxy, including 2387 new possible identifications and 134 known GCs. Their catalogue was later updated \citep{2024A&A...686A..42H} removing unbound stellar associations that were contaminating their sample, leaving 5647 cluster detections. \cite{2013A&A...558A..53K} used the Positions and Proper Motions Extended-L (PPMXL, \citealp{2010AJ....139.2440R}) and Two Micron All Sky Survey (2MASS; \citealp{2006AJ....131.1163S}) catalogues to identify and study around 3000 Galactic clusters, though their estimated kinematic parameters are considered to be unreliable for clusters beyond 2 kpc \citep{2013A&A...558A..53K}. \cite{2023ApJS..264....8H} used unsupervised machine learning and Gaia EDR3 data to detect 1656 star clusters beyond 1.2 kpc in the Galactic disk, including one GC candidate. \cite{2023ApJS..265...12Q} employed clustering algorithms on Gaia DR3 data to detect 324 OCs within 500 pc, including 101 new identifications. Finally, the catalogue of \cite{1996AJ....112.1487H} is a living (i.e. dynamically updated) database of Galactic GCs, currently containing 147 objects and their physical parameters, while the catalogue of \cite{2021MNRAS.505.5978V} contains all clusters from \cite{1996AJ....112.1487H} and their kinematic properties, together with some new GC candidates. \cite{2023A&A...673A.114H} contains 51.6\% of the clusters from \cite{2013A&A...558A..53K}, 44.5\% from \cite{2023ApJS..264....8H}, 73.3\% from \cite{2023ApJS..265...12Q}, and 78.8\% from \cite{2021MNRAS.505.5978V}. Some clusters from \cite{2013A&A...558A..53K} that used IR data may be unreachable with Gaia due to extinction effects and thus not detected by \cite{2023A&A...673A.114H}. The compilation of the previous catalogues provides a total sample of $\sim$ 9000 star clusters that are used in the following analysis.

\section{Identifying new candidate CS-cluster pairs} \label{sec:res}

All CSs from the \cite{2021A&A...656A..51G} catalogue were initially cross-matched with star clusters from the aforementioned cluster catalogues using an initially relaxed radius of 10 arcmin. CSs located outside the paired cluster tidal radius were subsequently removed from our sample. In the cases of the \cite{2023ApJS..264....8H} and \cite{2023ApJS..265...12Q} cluster catalogues, where the clusters' catalogued radii are either missing or very large and no subsequent positional cuts could be made, we selected a conservative cross-match radius of 4 arcmin. For \cite{1996AJ....112.1487H} and \cite{2021MNRAS.505.5978V} GCs, radii were adopted from \cite{2011CQGra..28h5016W}. 

Based on these strictly positional criteria, we found 237 potential CS-cluster pairs from the catalogue of \cite{2023A&A...673A.114H}, 77 from \cite{2013A&A...558A..53K}, 15 from \cite{2023ApJS..264....8H}, 2 from \cite{2023ApJS..265...12Q}, and 5 from \cite{1996AJ....112.1487H} and \cite{2021MNRAS.505.5978V}. From these, 51 are common to the \cite{2023A&A...673A.114H} and \cite{2013A&A...558A..53K} catalogues. Among these, one pair is also common to \cite{2023ApJS..265...12Q} and three to \cite{1996AJ....112.1487H} and \cite{2021MNRAS.505.5978V}. Moreover, four pairs are common to \cite{2023A&A...673A.114H} and \cite{2023ApJS..264....8H}. The \cite{1996AJ....112.1487H} and \cite{2021MNRAS.505.5978V} catalogues contributed to the exact same matches. Removing duplicates, we find a total of 277 potential CS-cluster pairs. 

Consequently, we proceeded to apply different selection criteria
in our sample of 277 pairs based on the consistency of the Gaia DR3 CS and cluster kinematic parameters for each pair. For the GCs in our sample, their kinematic parameters were obtained from \cite{2019MNRAS.482.5138B} and \cite{2021MNRAS.505.5978V}. For kinematic parameters without assigned errors in the literature, we adopt a conservative (i.e., generous) uncertainty of 30\%. In the following calculations, and since Gaia formal standard errors can be underestimated \citep{2021A&A...649A...5F}, we adopt for each CS the median uncertainties of the Gaia kinematic parameters based on its $G$ magnitude interval, as defined in the Gaia DR3 documentation\footnote{\url{https://www.cosmos.esa.int/web/gaia/dr3}}, as an additional tolerance term. For each parameter, the CS and cluster measurement errors are combined following standard error propagation, and the adopted tolerance term is added linearly to the resulting value to account for the documented \citesee{2021A&A...649A...5F} performance limits that could affect the reliability of our selection criteria. Although this approach could result in some false positives surviving our individual selection cuts, our primary goal is to exclude clear outliers with false positives expected to be rejected in subsequent steps. 

There are only two CSs in our sample with Gaia radial velocities, both potentially paired with OCs. Open cluster velocity dispersions are typically low ($\leq$ 1 \kms; \citealp{2000ASPC..198..517M}), so it is expected that in the case of true physical association the CS and cluster radial velocities should sufficiently match. Taking into account the typical cluster velocity dispersions, the assigned standard errors and the additional tolerance term defined above, we find that CS and cluster radial velocities of one of our pairs do not match, and thus we remove it from our sample, which now contains 276 pairs.

A total of 201 CSs in our sample have Gaia parallax and proper motion in right ascension and declination (pmRA and pmDEC) measurements. Thus, we apply further selection criteria based on these parameters. Taking into account the parallax assigned standard errors and corresponding tolerance term defined above, we found that for 50 pairs the CS and cluster parallaxes did not match (i.e., CS and cluster parallaxes differed by more than the quadratically combined parallax standard errors plus the adopted tolerance term), and we therefore removed them from our sample, leaving 226 pairs in total, including 151 with CS Gaia kinematic data. At this point, we also removed 15 pairs whose clusters are flagged as associations instead of bound clusters in the source catalogues (see Table~\ref*{tab:tabA1}), leaving 211 pairs (139 of which with CS Gaia kinematic data).

At distances equal to or larger than 1 kpc, internal cluster 2D proper motion dispersions are estimated to be around 0.4 mas/yr for OCs and 1 mas/yr for GCs \citep{2020A&A...633A..99C}. For OC distances (d) within 1 kpc, we conservatively estimate their 2D proper motion dispersions from \cite{2020A&A...633A..99C} (their Figure 1) as: 0.5 mas/yr for 0.6 $\leq$ d $<$ 1 kpc, 0.6 mas/yr for 0.5 $\leq$ d $<$ 0.6 kpc, and 2 mas/yr for d $<$ 0.5 kpc. There are no GCs in our sample with distances less than 1 kpc.

Assuming independent proper motion components and that each component contributes equally to the 2D dispersion, the 1D proper motion dispersions were calculated by dividing the 2D dispersions by $\sqrt{2}$, following $\sigma_{2D}^2 = 2\sigma_{1D}^2$ for isotropic motions. Considering the pmRA formal errors, the adopted pmRA tolerance term and the 1D cluster dispersions estimated above, we exclude from our sample 69 pairs based on inconsistent CS and cluster pmRAs, with 142 pairs left in our sample (70 with kinematic CS data). Similarly, we exclude 24 pairs based on pmDEC inconsistencies, which leaves us with 118 pairs, from which 46 have Gaia kinematic data. We also remove one pair because the Gaia Renormalised Unit Weight Error (RUWE) of its CS is larger than 1.4, rendering its kinematic data unreliable (see \citealt{2024A&A...688A...1C}, and references therein). 

Finally, we remove from our sample 32 pairs (see Table~\ref*{tab:tabB1}) with clusters younger than 50 Myr, since these are too young for MS stars with masses less than 8~\msun, which is the theoretical lower mass limit for core-collapse supernova formation \citeg{2013ApJ...765L..43I}, to have already left the MS \citep{2012MNRAS.427..127B} and evolved to a PN. We also remove one pair whose CS has been found as a misidentification by \cite{2025MNRAS.543.3035C}. Our final sample contains 84 CS-cluster pairs that cannot be excluded as possible true associations. From these, the 38 that have consistent kinematic properties, and thus a higher chance of being true physical associations, are listed in Table~\ref{tab:tab1}. The 46 pairs in our sample that lack CS kinematic data are presented in Table~\ref*{tab:tabC1} of the Appendix as a reference for future studies.

\begin{deluxetable*}{l c c c c c c c c c c c c}
\tablecaption{Candidate CS-cluster pairs. All clusters are OCs apart from NGC~6656 (CS\_ID=2312), which is a GC. } 
\label{tab:tab1}
\tabletypesize{\fontsize{7pt}{7pt}\selectfont}
\setlength{\tabcolsep}{1.0pt}
\tablewidth{0pt}
\tablehead{
\colhead{CS\_ID} & \colhead{PN} & \colhead{cs\_RAJ2000} & \colhead{cs\_DEJ2000} & \colhead{cs\_gr.\tablenotemark{1}} &
\colhead{Cluster} & \colhead{cl\_RAJ2000} & \colhead{cl\_DEJ2000} & \colhead{cl\_$R_t$(')\tablenotemark{2}} &
\colhead{cl\_orig.\tablenotemark{3}} & \colhead{separ.(')} & \colhead{KF\tablenotemark{4}} & \colhead{EF\tablenotemark{5}}
}
\startdata
206\tablenotemark{*} & MB4475 & 271.312 & -23.786 & B & HSC~97 & 271.286 & -23.803 & 7.51 & HR & 1.76 & S & EC \\
462 & M2-45 & 279.841 & -4.331 & B & HSC~278 & 279.956 & -4.369 & 12.96 & HR & 7.26 & M & EC \\
481 & Pa~34 & 276.314 & 0.035 & A & HSC~299 & 276.297 & -0.004 & 23.81 & HR & 2.52 & S & EI \\
494 & PHRJ1855-0116 & 283.821 & -1.282 & B & UBC~1049 & 283.863 & -1.251 & 8.28 & HR & 3.10 & M & EC \\
509 & PHRJ1900-0014 & 285.014 & -0.234 & A & HSC~324 & 285.002 & -0.306 & 6.50 & HR & 4.40 & W & EC \\
594 & IPHASXJ192510.5+113400 & 291.295 & 11.565 & B & Dolidze~35 & 291.372 & 11.637 & 14.10 & HR,K & 6.23 & M & EI \\
827 & KKR62 & 322.688 & 52.697 & B & Teutsch~17 & 322.753 & 52.835 & 12.08 & HR & 8.60 & W & EI \\
983 & Pu~2 & 85.642 & 36.152 & A & OC~0292 & 85.652 & 36.190 & 58.10 & HR & 2.31 & W & EI \\
992\tablenotemark{\S} & IPHASXJ055226.2+323724 & 88.109 & 32.624 & A & NGC~2099 & 88.080 & 32.541 & 30.42 & HR,K & 5.18 & S & EC \\
1022 & Abell~14 & 92.786 & 11.779 & B & HSC~1571 & 92.782 & 11.774 & 1.32 & HR,K & 0.38 & S & EI \\
1129 & NGC~2452 & 116.859 & -27.335 & A & NGC~2453 & 116.902 & -27.196 & 18.36 & HR & 8.65 & M & EI \\
1138 & Dr~2 & 115.229 & -33.557 & A & Bochum~15 & 115.072 & -33.551 & 31.95 & HR & 7.89 & W & EC \\
1174 & NGC~2818 & 139.006 & -36.627 & B & NGC~2818 & 139.042 & -36.621 & 15.30 & HR,K & 1.74 & M & EI \\
1213 & PHRJ0934-5223 & 143.573 & -52.389 & B & Teutsch~66 & 143.376 & -52.381 & 8.40 & HR & 7.24 & S & EC \\
1317\tablenotemark{\S} & Hf~69 & 175.405 & -62.482 & A & Lynga~15 & 175.488 & -62.494 & 15.62 & HR,K & 2.38 & S & EI \\
1330 & PHRJ1203-6403 & 180.803 & -64.062 & B & Juchert~13 & 180.434 & -64.101 & 15.03 & HR & 9.95 & M & EI \\
1359 & MGE301.8797-00.1036 & 190.545 & -62.957 & A & NGC~4609 & 190.567 & -62.993 & 37.76 & HR,K & 2.23 & S & EI \\
1424 & PHRJ1356-6139 & 209.187 & -61.663 & A & HSC~2622 & 209.186 & -61.614 & 4.41 & HR,K & 2.93 & M & EI \\
1529 & MPAJ1530-5801 & 232.712 & -58.025 & A & UBC~1537 & 232.812 & -57.984 & 13.22 & HR & 4.03 & M & EI \\
1541 & Hen~2-133 & 235.495 & -56.607 & A & Lynga~5 & 235.460 & -56.669 & 12.79 & HR,K & 3.90 & M & EI \\
1556 & vBe~3 & 238.246 & -56.407 & A & NGC~5999 & 238.044 & -56.480 & 13.36 & HR & 7.97 & S & EI \\
1675 & PHRJ1632-4549 & 248.237 & -45.825 & A & HSC~2801 & 248.419 & -45.813 & 8.19 & HR & 7.64 & S & EC \\
2085\tablenotemark{*} & JaSt~84 & 268.191 & -28.811 & C & HSC~10 & 268.226 & -28.797 & 6.59 & HR & 2.02 & W & EI \\
2258 & PPAJ1756-2311 & 269.138 & -23.197 & C & FSR~0022 & 269.107 & -23.192 & 7.50 & K & 1.74 & S & EC \\
2312 & GJJC~1 & 279.095 & -23.922 & C & NGC~6656 & 279.106 & -23.953 & 12.02 & HR,K,Har & 1.95 & S & EC \\
2387 & PHRJ1844-0452 & 281.069 & -4.881 & C & Andrews-Lindsay~5 & 281.078 & -4.930 & 6.97 & HR,K & 2.99 & S & EI \\
2472 & IPHASJ194727.51+241502.1 & 296.865 & 24.251 & C & Teutsch~7 & 296.934 & 24.261 & 4.12 & HR,K & 3.86 & M & EC \\
2478 & Pa~138 & 297.639 & 27.353 & C & CWNU~1753 & 297.727 & 27.432 & 44.38 & HR & 6.69 & M & EI \\
2479 & K~3-48 & 298.038 & 27.309 & C & UBC~130 & 298.061 & 27.449 & 13.57 & HR & 8.51 & S & EC \\
2507 & K~4-55 & 311.290 & 44.656 & C & HSC~659 & 311.235 & 44.621 & 5.86 & HR & 3.17 & M & EC \\
2535 & LBN157.08-03.61 & 65.177 & 44.943 & C & Berkeley~11 & 65.121 & 44.924 & 18.22 & HR,K & 2.64 & S & EI \\
2546 & IPHASXJ062937.8+065220 & 97.408 & 6.872 & C & NGC~2236 & 97.411 & 6.832 & 16.11 & HR,K & 2.41 & S & EI \\
2563 & PHRJ0753-2803 & 118.290 & -28.053 & C & ASCC~43 & 118.267 & -28.175 & 12.60 & K & 7.42 & W & EI \\
2567 & WRAY~17-15 & 123.341 & -33.017 & C & CWNU~1562 & 123.310 & -33.051 & - & He & 2.57 & S & EI \\
2591 & Hf~38 & 163.647 & -59.163 & C & CWNU~2232 & 163.504 & -59.188 & 16.86 & HR & 4.64 & M & EC \\
2605 & PHRJ1213-6344 & 183.442 & -63.750 & C & Theia~7490 & 183.352 & -63.757 & 69.42 & HR & 2.44 & M & EI \\
2652 & MPAJ1559-5552 & 239.796 & -55.872 & C & Theia~3118 & 239.805 & -55.870 & 13.87 & HR & 0.32 & S & EC \\
2821 & BlD & 266.513 & -31.060 & C & HSC~2974 & 266.500 & -31.028 & 8.92 & HR & 2.04 & W & EI \\
\enddata

\tablecomments{\scriptsize
(1) CS group from \GStwentyone;
(2) Cluster tidal radius;
(3) Source catalogue for cluster data. HR-\HRtwentythree, K–\Kharthirteen, He-\cite{2023ApJS..264....8H}, Har-\cite{1996AJ....112.1487H}. When a cluster is identified in more than one catalogue, employed data are prioritized in the indicated order; 
(4) Kinematic consistency flag: S- strong associations, M- moderate associations, W- weak associations; 
(5) Extinction consistency flag: EC- Compatible associations based on agreement of cluster and PN interstellar extinctions;
EI- Cluster and PN interstellar extinctions are incompatible;
(*) CS present NIR excess \Mintwentyfive;
(\S) These are the only CSs from our candidate pairs that are also in the sample of \cite{2025MNRAS.543.3035C} where they are found to be correct CS identifications (their flag A). 
}

\end{deluxetable*}

To assess the kinematic consistency of our pairs and flag possible false positives contaminating our sample, for each pair we computed a $\chi^2$-like kinematic score $K$ with an approximate three degrees of freedom, ignoring covariances and assuming Gaussian astrometric uncertainties \citeg{2018A&A...616A...2L}, as:

\begin{equation}
K =
\frac{(\Delta \mu_{\alpha*})^{2}}{\sigma_{\mu_{\alpha*}}^{2}} +
\frac{(\Delta \mu_{\delta})^{2}}{\sigma_{\mu_{\delta}}^{2}} +
\frac{(\Delta \varpi)^{2}}{\sigma_{\varpi}^{2}} .
\end{equation}

\noindent
where $\Delta \mu_{\alpha*}$, $\Delta \mu_{\delta}$, and $\Delta \varpi$ denote the pmRA, pmDEC and parallax offsets between the CS and the cluster respectively, while $\sigma_{\mu_{\alpha*}}^{2}$, $\sigma_{\mu_{\delta}}^{2}$, and $\sigma_{\varpi}^{2}$ are the effective variances for each parameter. The effective variances are estimated by adding in quadrature the formal CS and cluster measurement errors and, for proper motions, also including the intrinsic cluster dispersion in quadrature.

Our final sample does not include any CS with Gaia high-precision radial velocity data and thus radial velocities have not been taken into account. The corresponding PNe radial velocities were examined when available \citep{1998A&AS..132...13D,2025A&A...696A.146F}. Analyzing cluster membership requires precise systemic nebular radial velocities, whose measurement is a challenging task usually requiring multiple high-resolution observations obtained at symmetric pointings all around the PN. Thus, and due to the fact that available literature PN radial velocities originate from heterogeneous datasets of varying quality and resolution, we did not use them in our selection criteria, since this could lead to the rejection of genuine pairs \citeg{2025A&A...696A.146F}. For reference, only six PNe in our sample of 38 pairs (CS\_IDs: 983, 1129, 1174, 2312, 2591, 2821) have literature radial velocities, and of these only GJJC~1 (CS\_ID: 2312) is included in our final eight best candidates (see Section \ref{sec:dis}).

Based on the distribution of $K$ in our sample (0.056 $\leq$ $K$ $\leq$  13.99), we flag our pairs based on their kinematic consistency (with thresholds corresponding approximately to the $1\sigma$ and $2\sigma$ levels for a $\chi^2$ distribution with three degrees of freedom) as strong (S) associations ($K$ $\leq$ 3.51, 18 pairs) with an excellent kinematic consistency, moderate (M) associations (3.53 $<$ $K$ $\leq$ 7.82, 14 pairs) with a moderate kinematic consistency, and weak (W) associations ($K$ $>$ 7.81, 7 pairs) with relatively large kinematic offsets. The purpose of the aforementioned flags is to indicate kinematic consistency in our following analysis and does not imply an exclusion criterion.

Recent NIR studies have examined the sample of \cite{2021A&A...656A..51G} to identify CSs that present NIR excess and eclipsing binary candidates \citep{2025A&A...702A..79M}. By cross-matching their findings with our sample, we find that two CSs from our candidate CS-cluster pairs present NIR excess that is indicative of the presence of a binary cool companion and/or circumstellar disk \citep{2025A&A...702A..79M}. These candidates (CS\_ID: 206, 2085) are indicated in Table \ref{tab:tab1} with an asterisk. No eclipsing binary candidates have been found in our sample.

\section{Extinction and evolutionary constraints} \label{sec:ext}

To further clean our sample of candidate pairs, we proceed by examining our kinematically compatible pairs based on the consistency of their CS and cluster interstellar extinctions. Due to the lack of an homogeneous sample of total extinction determinations for the majority of the PNe in our sample, we adopt the CSs visual interstellar extinctions from \cite{2021A&A...656A..51G}, which are based on dust maps. Since their extinction values have no assigned formal errors, we assume a relaxed error threshold of 30\% to avoid rejecting genuine associations due to underestimated uncertainties. The adopted CSs extinctions do not account for the nebular internal extinction from local PN dust, which may cause an inconsistency with the clusters' extinctions, and thus are more appropriate for their direct comparison and our purposes. The cluster extinction values and their formal errors have been obtained from the cluster source catalogues. The adopted PN and cluster extinction values along with their parallaxes are presented in Table \ref{tab:tab2}. We define as pairs with compatible CS and cluster extinctions those with $|A_{V,\mathrm{cl}} - A_{V,\mathrm{cs}}| \leq A_{V,\mathrm{err}}$, where $A_{V,\mathrm{err}}$ is the CS and cluster assigned errors added in quadrature. The 15 extinction-compatible (EC) and the 23 extinction-incompatible (EI) pairs are flagged accordingly in Table \ref{tab:tab1}.

\begin{table}[ht!]
\centering
\scriptsize
\setlength{\tabcolsep}{5pt} 
\caption{The Gaia CS and literature cluster parallaxes, and their adopted extinctions for our sample of 38 candidate pairs.}
\label{tab:tab2}
\begin{tabular}{ccccc}
\hline
CS\_ID & cs\_$\varpi$ (mas)\tablenotemark{1} & cl\_$\varpi$ (mas) & pn\_$A_{V}$\tablenotemark{2} & cl\_$A_{V}$ \\
\hline
206 & $1.543 \pm 1.955$ & $0.326 \pm 0.004$ & 2.37 & $2.54^{+0.20}_{-0.26}$ \\
462 & $-0.073 \pm 0.702$ & $0.610 \pm 0.003$ & 2.73 & $2.81^{+0.75}_{-1.12}$ \\
481 & $0.401 \pm 0.290$ & $0.171 \pm 0.003$ & 2.39 & $4.80^{+0.76}_{-0.60}$ \\
494 & $2.638 \pm 1.261$ & $0.311 \pm 0.003$ & 3.19 & $3.39^{+0.19}_{-0.20}$ \\
509 & $-0.688 \pm 1.063$ & $0.308 \pm 0.005$ & 2.81 & $3.45^{+0.32}_{-0.31}$ \\
594 & $0.856 \pm 0.811$ & $0.414 \pm 0.005$ & 3.18 & $4.25^{+0.27}_{-0.20}$ \\
827 & $1.061 \pm 0.530$ & $0.296 \pm 0.005$ & 2.30 & $3.77^{+0.23}_{-0.21}$ \\
983 & $0.026 \pm 0.389$ & $0.575 \pm 0.003$ & 2.74 & $1.76^{+0.24}_{-0.26}$ \\
992 & $1.020 \pm 0.315$ & $0.673 \pm 0.001$ & 0.51 & $0.62^{+0.14}_{-0.17}$ \\
1022 & $0.218 \pm 0.029$ & $0.210 \pm 0.017$ & 1.17 & $1.83^{+0.23}_{-0.23}$ \\
1129 & $0.342 \pm 0.079$ & $0.219 \pm 0.003$ & 0.81 & $1.22^{+0.19}_{-0.19}$ \\
1138 & $0.058 \pm 0.540$ & $0.257 \pm 0.003$ & 2.21 & $1.58^{+0.22}_{-0.23}$ \\
1174 & $0.032 \pm 0.211$ & $0.300 \pm 0.002$ & 0.43 & $0.19^{+0.08}_{-0.13}$ \\
1213 & $0.474 \pm 0.528$ & $0.215 \pm 0.004$ & 3.91 & $3.08^{+0.26}_{-0.20}$ \\
1317 & $0.430 \pm 0.248$ & $0.563 \pm 0.003$ & 15.92 & $0.82^{+0.20}_{-0.15}$ \\
1330 & $-0.126 \pm 0.266$ & $0.316 \pm 0.003$ & 5.31 & $2.21^{+0.30}_{-0.18}$ \\
1359 & $0.325 \pm 0.700$ & $0.688 \pm 0.003$ & 12.40 & $1.04^{+0.15}_{-0.20}$ \\
1424 & $-0.260 \pm 1.465$ & $0.290 \pm 0.003$ & 26.18 & $5.34^{+0.41}_{-0.30}$ \\
1529 & $0.241 \pm 0.473$ & $0.450 \pm 0.004$ & 7.01 & $3.24^{+0.25}_{-0.18}$ \\
1541 & $1.369 \pm 0.605$ & $0.336 \pm 0.003$ & 9.03 & $4.25^{+0.16}_{-0.22}$ \\
1556 & $0.462 \pm 0.617$ & $0.349 \pm 0.002$ & 2.77 & $1.79^{+0.16}_{-0.21}$ \\
1675 & $0.763 \pm 0.451$ & $0.200 \pm 0.005$ & 5.03 & $4.81^{+0.31}_{-0.35}$ \\
2085 & $2.866 \pm 1.144$ & $0.351 \pm 0.003$ & 1.22 & $2.45^{+0.32}_{-0.36}$ \\
2258 & $0.669 \pm 1.205$ & $0.230 \pm 0.069$ & 4.57 & $5.42^{+1.63}_{-1.63}$ \\
2312 & $0.330 \pm 0.042$ & $0.284 \pm 0.001$ & 0.94 & $1.07^{+0.32}_{-0.32}$ \\
2387 & $0.443 \pm 0.698$ & $0.213 \pm 0.004$ & 1.67 & $4.42^{+0.24}_{-0.28}$ \\
2472 & $0.837 \pm 0.775$ & $0.166 \pm 0.008$ & 3.81 & $4.85^{+0.71}_{-0.74}$ \\
2478 & $-1.178 \pm 0.740$ & $0.449 \pm 0.003$ & 3.93 & $1.68^{+0.24}_{-0.16}$ \\
2479 & $0.475 \pm 0.179$ & $0.395 \pm 0.002$ & 1.66 & $2.00^{+0.19}_{-0.19}$ \\
2507 & $-0.326 \pm 0.589$ & $0.381 \pm 0.003$ & 4.12 & $3.60^{+0.45}_{-0.43}$ \\
2535 & $0.321 \pm 0.049$ & $0.357 \pm 0.005$ & 1.82 & $2.71^{+0.16}_{-0.23}$ \\
2546 & $0.493 \pm 0.091$ & $0.348 \pm 0.003$ & 0.74 & $1.38^{+0.16}_{-0.19}$ \\
2563 & $0.322 \pm 0.542$ & $0.995 \pm 0.299$ & 0.86 & $0.19^{+0.06}_{-0.06}$ \\
2567 & $-1.543 \pm 1.664$ & $0.490 \pm 0.030$ & 1.47 & $0.95^{+0.05}_{-0.05}$ \\
2591 & $-0.896 \pm 0.532$ & $0.207 \pm 0.003$ & 2.94 & $2.38^{+0.26}_{-0.21}$ \\
2605 & $-0.844 \pm 1.187$ & $0.521 \pm 0.004$ & 2.44 & $0.91^{+0.21}_{-0.21}$ \\
2652 & $0.412 \pm 0.090$ & $0.329 \pm 0.003$ & 3.13 & $2.20^{+0.18}_{-0.21}$ \\
2821 & $0.748 \pm 0.164$ & $0.515 \pm 0.008$ & 1.47 & $3.03^{+0.18}_{-0.30}$ \\
\hline
\end{tabular}
\footnotesize

\tablecomments{\scriptsize
(1) Negative Gaia parallaxes for some CSs result from the statistical nature of Gaia parallax measurements. Although these measurements are not reliable as absolute values, CS parallaxes were used here to remove obvious outliers during candidate CS-cluster pair selection. In the subsequent analysis of the most promising candidates, positive CS Gaia parallaxes were converted into distances, while cluster distances were adopted for the cases with negative CS parallaxes;
(2) CS extinction from \GStwentyone.
}
\end{table}

Our final kinematically and extinction-consistent candidate CS-cluster sample (hereafter referred to as sample A*) contains 15 pairs (see Table \ref{tab:tab3}). Eight of these pairs have strong kinematic consistency (S flag), two of which have CSs in group A of \cite{2021A&A...656A..51G}, two in group B and four in group C. Five pairs have moderate kinematic consistency (M flag), with two CSs also in group B of \cite{2021A&A...656A..51G} and three in group C. Finally, two pairs have weak kinematic consistency (W flag), with both CSs belonging to group A of \cite{2021A&A...656A..51G}. None of the CSs of our A* sample have assigned evolutionary parameters in \cite{2021A&A...656A..51G}.

Following \cite{2021A&A...656A..51G} and \cite{2025MNRAS.543.3035C}, we further examine our A* sample to assess the consistency of the CSs with evolutionary models. We adopt the temperature $T_{\rm eff}$ and luminosity $L/L_{\odot}$ estimates for the CSs of GJJC~1 (designated also as IRAS 18333-2357) and IPHASXJ055226.2+323724 from \cite{2020A&A...640A..10W} and \cite{2022ApJ...935L..35F}, respectively, which are the only in our A* sample with $T_{\rm eff}$ and $L/L_{\odot}$ data in the literature. Since \cite{2020A&A...640A..10W} do not assign errors for their reported values, we assume again a 30\% uncertainty threshold. Apart from the CS of GJJC~1 \citesee{2017ApJ...836...93J}, there are no available UV data for any other CS in our A* sample. Four PNe (M~2-45, PHRJ0934-5223, K~4-55 and Hf~38) have literature dereddened~$H_\alpha$~fluxes \citep{2013MNRAS.431....2F} and/or surface brightness measurements \citep{2016MNRAS.455.1459F}. In the case of PN K4-55, for which only an extinction-corrected $H_\alpha$~surface brightness is available, we convert to dereddened~$H_\alpha$~flux adopting the PN radius from \cite{2016JPhCS.728c2008P}. The $H_\alpha$~fluxes were then converted to $H_\beta$~fluxes adopting the standard Balmer decrement of $H_{\alpha}/H_{\beta} = 2.86$. 

Consequently, we convert the CS $G$ Gaia magnitudes to visual $V$ Johnson magnitudes (required for the subsequent calculations of the CSs Zanstra temperatures and bolometric corrections) using the standard recipe from the Gaia documentation\footnote{\url{https://gea.esac.esa.int/archive/documentation/GDR3/Data_processing/chap_cu5pho/cu5pho_sec_photSystem/cu5pho_ssec_photRelations.html}}, which is generally reliable for hot stars \citep{2021A&A...656A..51G}. For CSs with missing Gaia colours (PPAJ1756-2311, Hf~38 and MPAJ1559-5552), we assume a $V = G + 0.7$ and adopt an uncertainty of 0.7 mag to account for the simplified colour conversion. Visual magnitudes have been corrected for extinction using the visual extinction values from \cite{2021A&A...656A..51G}. For the four PNe with available dereddened~$H_\beta$~fluxes (see above), we used their extinction-corrected CS $V$ magnitudes for the estimation of their Zanstra HI effective temperatures \citep{1931ZA......2....1Z}. For the estimated Zanstra temperatures we assume an error of 30\% (see, \citealp{2019MNRAS.484.3078F}, and references therein). Zanstra HI temperatures assume optical thickness to the H-ionizing radiation, which is unlikely for some of the more extended PNe in our sample. Due to the lack of He II fluxes that would allow the estimation of their He II Zanstra temperatures, which are more reliable for extended PNe \citep{2025MNRAS.543.3035C}, we assumed that all four PNe are optically thick to H-ionizing radiation. For PNe in our A* sample with neither $T_{\rm eff}$ nor $H_\alpha$~fluxes in the literature, we assumed a $T_{\rm eff}=100^{+100}_{-50}$ kK to account for a wide range of PNe temperatures typical for CSs \citep{2016A&A...588A..25M, 2025MNRAS.543.3035C}. In all such cases, and due to the lack of literature data for their total PN extinctions, the \cite{2021A&A...656A..51G} extinctions, based on dust maps, were used instead, which do not account for the nebular internal extinction and may underestimate the total extinction value, especially for young and compact PNe.

Subsequently, we estimated the CS distances from their Gaia parallaxes and assigned uncertainties using standard error propagation. Four CSs in our A* sample (M~2-45, PHRJ1900-0014, K~4-55 and Hf~38) have negative Gaia parallaxes, preventing their direct use, so we adopt the parallax of their associated clusters. The CS absolute visual magnitudes $M_V$ were then computed from their extinction-corrected $V$ magnitudes and distances. The bolometric correction, for each CS, was estimated as $\mathrm{BC} = 27.66 - 6.84 \cdot \log(T_{\rm eff})$ \citep{1996ApJ...460..914V}. After applying the corresponding bolometric correction to the estimated absolute magnitudes, we derived the CS luminosities as $L/L_{\odot} = 10^{-0.4 (M_{\rm bol} - M_{{\rm bol},\odot})}$, where $M_{{\rm bol},\odot} = 4.74$ \citep{1976asqu.book.....A}. All errors have been calculated through standard error propagation.

\begin{figure*}[]
\plotone{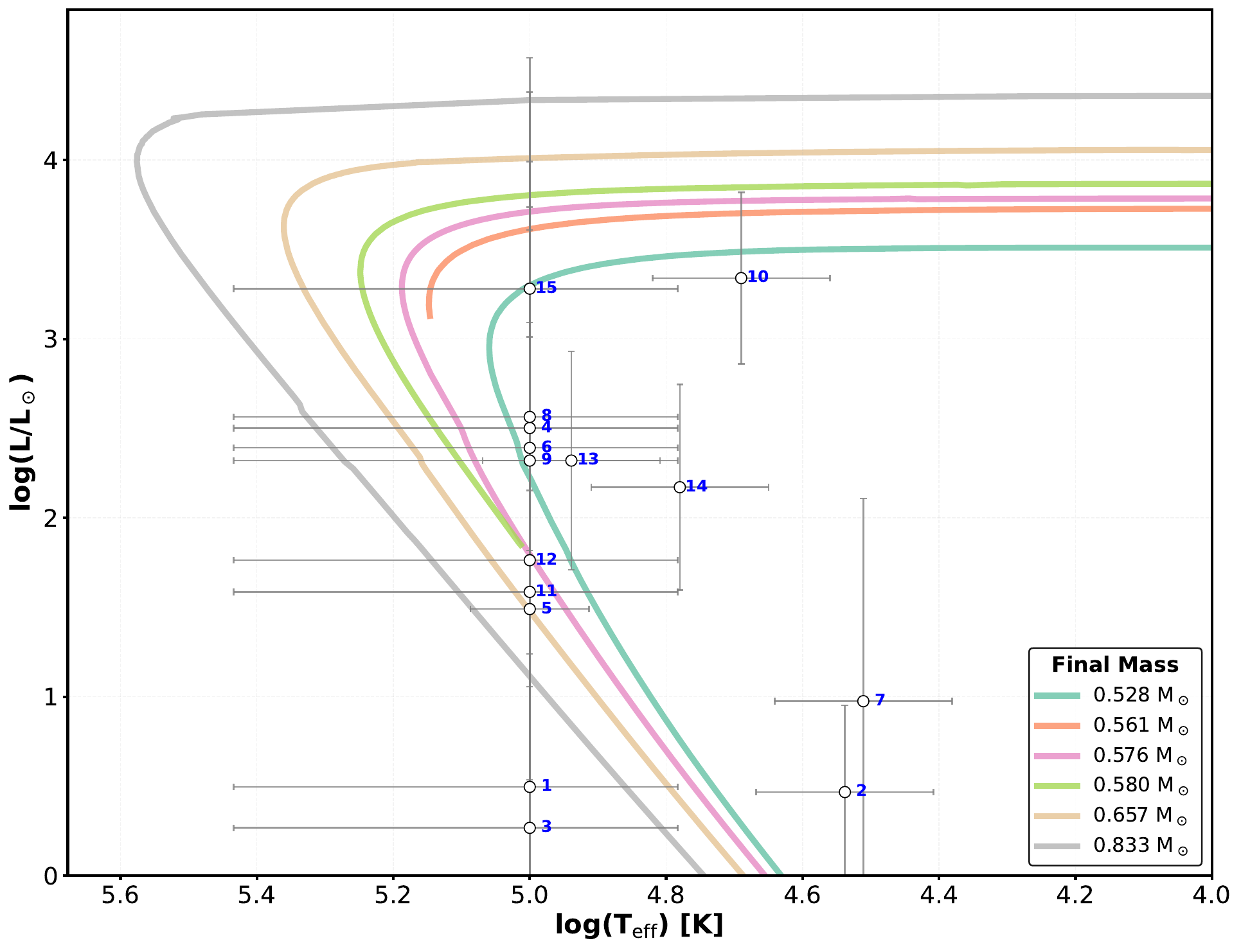}
\caption{The 15 candidates in our A* sample and their errors overplotted along the \cite{2016A&A...588A..25M} H-burning solar post-AGB evolutionary tracks (solid coloured lines). Individual points indicate the position of our CSs in the diagram, with the following labels: 
1 — MB4475 (CS\_ID = 206); 2 — M~2-45 (CS\_ID = 462); 3 — PHRJ1855-0116 (CS\_ID = 494); 4 — PHRJ1900-0014 (CS\_ID = 509); 5 — IPHASXJ055226.2+323724 (CS\_ID = 992);
6 — Dr~2 (CS\_ID = 1138); 7 — PHRJ0934-5223 (CS\_ID = 1213); 8 — PHRJ1632-4549 (CS\_ID = 1675); 9 — PPAJ1756-2311 (CS\_ID = 2258);
10 — GJJC~1 (CS\_ID = 2312); 11 — IPHASJ194727.51+241502.1 (CS\_ID = 2472); 12 — K~3-48 (CS\_ID = 2479); 13 — K~4-55 (CS\_ID = 2507);
14 — Hf~38 (CS\_ID = 2591); 15 — MPAJ1559-5552 (CS\_ID = 2652).
\label{fig:fig1}}
\end{figure*}

All CS $T_{\rm eff}$ and $L/L_{\odot}$ points and their errors were plotted along the \cite{2016A&A...588A..25M} H-burning post-AGB evolutionary tracks for all available metallicities, as shown in Figure \ref{fig:fig1}. We find that, based on their positions along the tracks, irrespective of metallicity, three of the sources in our A* sample (the reported CSs of M~2-45, PHRJ0934-5223 and Hf~38) are unlikely to be post-AGB stars, and so we do not include them in the following analysis. Both the CSs of MB4475 and PHRJ1855-0116 fall outside the area covered by the tracks and towards higher masses, but since there are no available tracks for relatively high-mass stars, we keep them in our A* sample.

For the 12 CSs left in our A* sample we estimate their final masses and their errors by interpolating their $T_{\rm eff}$ vs $L/L_{\odot}$ points along the corresponding evolutionary tracks. For clusters with no metallicity estimates in the literature, we assumed a solar value. We use the solar metallicity (Z=0.02) post-AGB evolutionary tracks for all CSs apart from GJJC~1, for which we use the subsolar metallicity (Z=0.0001) tracks, compatible with the literature metallicity of its associated cluster. The derived masses and their errors were truncated to reflect physically plausible CS masses. Finally, we estimate the CSs initial masses using cluster theoretical isochrones \citep{2012MNRAS.427..127B}, taking into account the time that has passed since the stars left the MS. The estimated initial-mass errors reflect the cluster age and metallicity uncertainties. The estimated final mass errors for all CSs, apart from those of PNe IPHASXJ055226.2+323724 (CS\_ID=992), GJJC~1 (CS\_ID=2312) and K~4-55 (CS\_ID=2507), whose $T_{\rm eff}$ is based on literature data, are large, reflecting the wide CS $T_{\rm eff}$ range adopted in our calculations. Our results for all pairs in sample A* are presented in Table \ref{tab:tab3}. 

\begin{deluxetable*}{lccccccc}
\tablecaption{Derived and literature parameters for the best candidate CS-cluster pairs (sample A*)\label{tab:tab3}}
\tablehead{
\colhead{CS\_ID} &
\colhead{cs\_$\log(T_{\rm eff}/{\rm K})$\tablenotemark{1}} &
\colhead{cs\_$\log(L/L_\odot)$\tablenotemark{1}} &
\colhead{cl\_$\log({\rm Age/yr})$\tablenotemark{2}} &
\colhead{cl\_[Fe/X]\tablenotemark{3}} &
\colhead{cs\_$M_{\rm init}\,(M_\odot)$} &
\colhead{cs\_$M_{\rm fin}\,(M_\odot)$} &
\colhead{Notes\tablenotemark{4}}
}
\startdata
206\tablenotemark{\S}  & $5.00^{+0.43}_{-0.22}$ & $0.50^{+1.66}_{-1.30}$ & $8.18^{+0.22}_{-0.22}$ & $0.10\pm0.13$ & $4.35^{+1.00}_{-0.82}$ & $0.86^{+0.24}_{-0.01}$ & consistent \\
462  & $4.54^{+0.13}_{-0.13}$ & $0.47^{+0.48}_{-0.48}$ & $9.77^{+0.23}_{-0.39}$ & $0.00$ &  &  & not a CS \\
494\tablenotemark{\S}  & $5.00^{+0.43}_{-0.22}$ & $0.27^{+1.32}_{-0.82}$ & $7.73^{+0.18}_{-0.22}$ & $0.00$ & $6.67^{+1.33}_{-1.13}$ & $0.90^{+0.20}_{-0.02}$ & consistent \\
509\tablenotemark{\S}  & $5.00^{+0.43}_{-0.22}$ & $2.50^{+1.24}_{-0.68}$ & $9.01^{+0.31}_{-0.27}$ & $0.00$ & $2.10^{+0.52}_{-0.46}$ & $0.51^{+0.22}_{-0.01}$ & consistent \\
992\tablenotemark{*}\tablenotemark{\S} & $5.00^{+0.09}_{-0.09}$ & $1.49^{+0.25}_{-0.25}$ & $8.69^{+0.18}_{-0.18}$ & $0.09\pm0.14$ & $2.76^{+0.50}_{-0.42}$ & $0.65^{+0.08}_{-0.05}$ & consistent \\
1138 & $5.00^{+0.43}_{-0.22}$ & $2.39^{+1.22}_{-0.65}$ & $7.77^{+0.22}_{-0.29}$ & $0.00$ & $6.40^{+1.60}_{-1.28}$ & $0.52^{+0.22}_{-0.02}$ & inconsistent \\
1213 & $4.51^{+0.13}_{-0.13}$ & $0.98^{+1.13}_{-1.13}$ & $8.38^{+0.28}_{-0.15}$ & $0.44\pm0.15$ &  &  & not a CS \\
1675\tablenotemark{\S} & $5.00^{+0.43}_{-0.22}$ & $2.56^{+1.43}_{-0.99}$ & $8.20^{+0.30}_{-0.26}$ & $0.00$ & $4.20^{+1.18}_{-1.60}$ & $0.51^{+0.22}_{-0.01}$ & consistent \\
2258\tablenotemark{\S} & $5.00^{+0.43}_{-0.22}$ & $2.32^{+2.06}_{-1.78}$ & $9.65^{+0.08}_{-0.10}$ & $0.00$ & $1.20^{+0.11}_{-0.05}$ & $0.52^{+0.23}_{-0.02}$ & consistent \\
2312\tablenotemark{\S} & $4.69^{+0.13}_{-0.13}$ & $3.34^{+0.48}_{-0.48}$ & $10.10^{+0.08}_{-0.10}$ & $-1.55\pm0.46$ & $0.78^{+0.02}_{-0.04}$ & $0.50^{+0.06}_{-0.00}$ & borderline \\
2472\tablenotemark{\S} & $5.00^{+0.43}_{-0.22}$ & $1.59^{+1.51}_{-1.10}$ & $9.11^{+0.48}_{-0.38}$ & $0.13\pm0.04$ & $1.97^{+0.74}_{-0.73}$ & $0.62^{+0.31}_{-0.12}$ & consistent \\
2479 & $5.00^{+0.43}_{-0.22}$ & $1.76^{+1.25}_{-0.71}$ & $7.94^{+0.19}_{-0.22}$ & $0.00$ & $5.38^{+1.37}_{-0.89}$ & $0.60^{+0.28}_{-0.10}$ & borderline \\
2507 & $4.94^{+0.13}_{-0.13}$ & $2.32^{+0.61}_{-0.61}$ & $7.98^{+0.59}_{-0.57}$ & $0.00$ & $5.17^{+2.83}_{-2.14}$ & $0.50^{+0.04}_{-0.00}$ & inconsistent \\
2591 & $4.78^{+0.13}_{-0.13}$ & $2.17^{+0.57}_{-0.57}$ & $7.73^{+0.27}_{-0.23}$ & $0.00$ &  &  & not a CS \\
2652 & $5.00^{+0.43}_{-0.22}$ & $3.28^{+1.29}_{-0.78}$ & $7.73^{+0.21}_{-0.22}$ & $0.00$ & $6.67^{+1.33}_{-1.34}$ & $0.53^{+0.41}_{-0.03}$ & borderline \\
\enddata
\tablecomments{\scriptsize
(1) For the CSs of IPHASXJ055226.2+323724 (CS\_ID=992) and GJJC~1 (CS\_ID=2312), $T_{\rm eff}$ and $L$ were obtained from \cite{2022ApJ...935L..35F} and \Weidtwenty, respectively. The temperatures of the CSs of M~2-45 (CS\_ID=462), PHRJ0934-5223 (CS\_ID=1213), K4-55 (CS\_ID=2507) and Hf~38 (CS\_ID=2591) are the Zanstra temperatures estimated using the PN dereddend $H_\alpha$~ fluxes \cite{2013MNRAS.431....2F, 2016MNRAS.455.1459F}, and have an assigned uncertainty of 30\%. For the rest of our objects we assume a temperature of $100^{+100}_{-50}$ kK to account for a wide range of possible CS temperatures;
(2) Obtained from \HRtwentythree, apart from the age of the OC FSR~0022 (CS\_ID=2258), which was obtained from \Kharthirteen; 
(3) Metallicity for clusters HSC\_97 (CS\_ID=205) and Teutsch\_66 (CS\_ID=1213) obtained from \cite{2021MNRAS.504..356D}, for clusters NGC~2099 (CS\_ID=992) and NGC~6656 (CS\_ID=2312) from \Kharthirteen~and for cluster Teutsch~7 (CS\_ID=2472) from \Carninteen. For the rest of the clusters we assume a solar metallicity; 
(4) Notes on cluster-CS status based on consistency of their initial and final masses. CSs whose $T_{\rm eff}$ and $L$ are inconsistent with evolved stars are flagged as not CSs;
(*) This is the only CS from our A* sample that is also in the sample of \cite{2025MNRAS.543.3035C}, where it is found to be a correct CS identification (their flag A).  The corresponding PN is a confirmed member of the OC NGC~2099 \citep{2022ApJ...935L..35F}.
(\S) The eight most promising candidates from our sample.} 
\end{deluxetable*}

\section{Discussion} \label{sec:dis}

Although our selection criteria were deliberately made less restrictive to avoid rejecting true cluster-PN associations, some confirmed and/or reported pairs have not survived our cuts. The CSs of PN PHRJ1315-6555, which is a confirmed member of the OC ESO 96-SC04 \citep{2011MNRAS.413.1835P}, and PN JaFu 2 that is associated with the GC NGC~6441 \citep{1997AJ....114.2611J}, have been rejected based on our parallax selection criteria (i.e. the CS Gaia parallaxes were inconsistent to their respective clusters' parallaxes). Similarly, the CSs of PN BMPJ1613-5406, which is a confirmed member of the OC NGC~6067 \citep{2019NatAs...3..851F}, and PN JaFu~1, which is associated with the GC Palomar~6 \citep{1997AJ....114.2611J}, have been rejected based on our pmRA selection criteria (i.e. the CS Gaia pmRAs were inconsistent to their respective clusters' pmRAs). 

The CSs of PHRJ1315-6555, JaFu~1, and JaFu~2 are in the low-quality sample of \cite{2021A&A...656A..51G}, namely their catalogue C, which means that there is an elevated chance of these CSs being misidentifications. Moreover, parallaxes would be too small to be definitive in the case of distant objects, such as PHRJ1315-6555 and JaFu~2. \cite{2019MNRAS.484.3078F}, using HST data, identified a much fainter CS of PN PHRJ1315-6555 that is not accessible with Gaia, highlighting the limitations of the study of \cite{2021A&A...656A..51G} and our methodology for distant and faint CSs. Based on HST data, the proper motions of the CS of JaFu~2 have been found to agree with these of the GC NGC~6441 \citep{2020AJ....159..276B}, while the proper motions of the CS of JaFu~1 have been found to be inconsistent to the proper motions of the GC Palomar~6 \citep{2024AJ....168..160B}. The CS of PN BMPJ1613-5406 is in the high-quality sample of \cite{2021A&A...656A..51G} (their group A) and has been found to be the only blue star in the nebular field by deep FORS2 imaging \citep{2022Galax..10...44F}, though this CS identification could not be confirmed as reliable by \cite{2025MNRAS.543.3035C}. Moreover, the CS of PN PHRJ1724-3859, which is a confirmed member of the OC Trumpler 25 \citep{2026ApJ...996...90F}, has been rejected from our sample because it has been found to be a CS misidentification \citep{2025MNRAS.543.3035C}. For this PN, \cite{2026ApJ...996...90F} has identified, using VLT data, a fainter source, which is unreachable with Gaia, as the true CS. These instances show that our method, which relies on the reliability of the \cite{2021A&A...656A..51G} catalogue, is not exhaustive, and thus cannot rule out true associations. 

Moreover, although the PN NGC~2818 is a confirmed member of the OC NGC~2818A \cite{2025A&A...696A.146F}, in this work we find their extinctions to be incompatible. \cite{2023A&A...673A.114H}, which is our source catalogue for the physical parameters of the cluster NGC~2818A, found a cluster extinction of $A_V = 0.19^{+0.08}_{-0.13}$, while \cite{2025A&A...696A.146F}, using multiple literature data, estimated a cluster $A_V=0.51 \pm 0.25$, which is more consistent with the PN extinction value of 0.43 from \cite{2021A&A...656A..51G}. Therefore, our extinction flags should be treated with caution and not as definitive indicators of incompatible associations.

No CS from our A* sample is identified as part of a possible close or wide binary system \citep{2020A&A...644A.173G,2021A&A...656A..51G}. Inspecting the estimated initial and final masses of the CSs of our A* sample, we found that these are not consistent in the cases of the CSs of PNe Dr~2 (CS\_ID=1138) and K~4-55 (CS\_ID=2507). Based on stellar evolution models for single stars \citeg{2016A&A...588A..25M}, the derived final masses are too low for the massive progenitors that the age of their suspected host clusters predict. The extreme mass loss associated with interacting binary systems \citesee{2011ApJ...730...67B}, which are expected to be common among PN progenitors \citep{2009PASP..121..316D}, could explain such cases and add noise and uncertainty to the IFMR. Since there is currently no direct evidence of these two CSs to be part of a binary system, we infer that these two cluster-PN pairs are unlikely to be true associations assuming single star evolution, and we exclude them from our analysis, which assumes the absence of a binary companion. 

The initial and final masses of the CSs of PNe GJJC~1 (CS\_ID=2312), K~3-48 (CS\_ID=2479) and MPAJ1559-5552 (CS\_ID=2652) are borderline consistent with theoretical predictions assuming single star evolution. In the case of GJJC~1, the predicted progenitor mass is too low to produce a visible PN, though it could be explained as the result of a merger event (see below; \citealp{2017ApJ...836...93J}). For PNe K~3-48 and MPAJ1559-5552, the estimated initial and final CS masses present a small overlap within their large assigned errors according to stellar evolution models, and thus we cannot exclude them without more precise measurements. The CS masses of all other pairs are found to be consistent with single star evolution.

\subsection{Individual candidate cluster-PN pairs}\label{sub:ind}

We proceed to inspect the pairs whose CS masses were previously found to be consistent with single star evolution, including the dubious GJJC~1 case, separately. This allows us to evaluate in more detail their plausibility as genuine associations.

\subsubsection{PN MB4475 - OC HSC~97 (CS\_ID=206)}

MB4475 (candidate 1 in Figure~\ref{fig:fig1}) is a faint nebula, undetectable in the optical regime, with an angular diameter of 30 arcsec, currently classified as a possible PN \citep{2016JPhCS.728c2008P}. The nebula is visible in mid-IR data (MIPSGAL 24~$\mu$m regime; \citealp{2009PASP..121...76C}), having a round shape. High initial mass post-AGB stars evolve fast and their young, bright PN phase is therefore too short, having a low probability of being observed. On the other hand, their visibility timescale lasts longer as the more massive nebula continues expanding and remains detectable for longer due to its greater emission measure. As a result, the evolved PN phase of high-mass progenitors is more likely to be observed \citeg{2019NatAs...3..851F,2026ApJ...996...90F}. In the case of MB4475, both the estimated initial and final CS masses are relatively high, suggesting a higher probability of observing an evolved and therefore extended and faint PN, consistent with its observed characteristics. Its CS, in the group B of \cite{2021A&A...656A..51G}, presents IR excess indicative of the presence of a circumstellar disk and/or a binary cool companion (see Section 3.2 of \citealp{2025A&A...702A..79M}). Given an intermediate-mass progenitor as suggested by our study, and considering that the nebula is extremely faint in the optical but presents mid-IR emission, we suggest that a circumstellar disk surrounding a very evolved PN is more likely than a cool companion. More data are required to clarify the nature of this object and its possible association with the OC HSC~97.

\subsubsection{PN PHRJ1855-0116 - OC UBC~1049 (CS\_ID=494)}

PHRJ1855-0116 (candidate 3 in Figure~\ref{fig:fig1}) is a true, extended, bipolar PN with a Type I chemistry \citep{1983IAUS..103..233P} and an angular diameter of 47 arcsec  \citep{2016JPhCS.728c2008P}, typical of PNe associated with OCs \citeg{2026ApJ...996...90F}. The estimated high initial and final masses of its CS, which is included in the group B of \cite{2021A&A...656A..51G}, are consistent with the observed PN properties, rendering this possible cluster-PN pair an excellent candidate for follow up. 

\subsubsection{PN PHRJ1900-0014 - OC HSC~324 (CS\_ID=509)}

PHRJ1900-0014 (candidate 4 in Figure~\ref{fig:fig1}) is a true, elliptical, possible Type I PN with a relatively large angular diameter of 48 arcsec \citep{2016JPhCS.728c2008P}. The derived final mass of its CS (0.51~\msun), which is in the high-quality sample (group A) of \cite{2021A&A...656A..51G}, is relatively low for such an extended and evolved PN formed by a 2.1~\msun~progenitor, but the large assigned final mass errors could account for this discrepancy. Note that the CS-cluster pair has a relatively weak kinematic consistency (see Section \ref{sec:res}). Better data are needed both to constrain a more precise CS final mass and to confirm the association of the two objects.

\subsubsection{PN IPHASXJ055226.2+323724 - OC NGC~2099 (CS\_ID=992)}

IPHASXJ055226.2+323724 (candidate 5 in Figure~\ref{fig:fig1}) is a confirmed member of OC NGC~2099 and its CS (in the group A of \citealp{
2021A&A...656A..51G}) was independently identified by \cite{2022ApJ...935L..35F}, whose $T_{\rm eff}$ and $L$ estimates are adopted in this study. As expected, the derived initial and final CS masses are compatible with the literature values. 

\subsubsection{PN PHRJ1632-4549 - OC HSC~2801 (CS\_ID=1675)}

PHRJ1632-4549 (candidate 8 in Figure~\ref{fig:fig1}) is a likely, irregular, extended and possible Type I PN with an angular diameter of 62 arcsec \citep{2016JPhCS.728c2008P}. Its CS is in the high-quality sample of \cite{2021A&A...656A..51G} (their group A). The estimated CS final mass of 0.51~\msun~is too low for an intermediate-mass progenitor of 4.20~\msun, but still compatible within the large errors. The relatively high progenitor mass is consistent with an extended and Type I PN, as suggested by the observational nebular characteristics. More data are needed to confirm the PN nature of the nebula and to better constrain the CS final mass. 

\subsubsection{PN PPAJ1756-2311 - OC FSR~0022 (CS\_ID=2258)}

PPAJ1756-2311 (candidate 9 in Figure~\ref{fig:fig1}) is a true, round PN with weak He II emission and a diameter of 9 arcsec \citep{2016JPhCS.728c2008P}. Its CS is in the group C of \cite{2021A&A...656A..51G}. The estimated low CS final mass (0.52~\msun) is consistent with a progenitor mass of 1.20~\msun, though cluster PNe are usually extended and evolved from higher-mass stars \citeg{2026ApJ...996...90F}. The Gaia parallax (0.67 $\pm$ 1.20), which was used for the estimation of the CS distance, has a large assigned error. If this error is taken into account, a larger PN distance could explain its small angular size while still being compatible with the cluster distance of 4.34 kpc. A compatible higher-mass progenitor is consistent within the errors of our initial-mass estimates. More data are needed to clarify this issue. 

\subsubsection{PN GJJC~1 - GC NGC~6656 (CS\_ID=2312)}

The CS of GJJC~1 (candidate 10 in Figure~\ref{fig:fig1}), in the group C of \cite{
2021A&A...656A..51G}, has been independently identified by \cite{2020A&A...640A..10W}. GJJC~1 is reported as physically associated with GC NGC~6656 \citep{1989ApJ...338..862G}, with recent studies questioning the true nature of this nebula due to its peculiar characteristics, which could also suggest binary evolution \citep{2017ApJ...836...93J}. Adopting a cluster age of 13.1 $\pm$ 1.2 Gyr and a cluster metallicity of [Fe/H]=-1.7, and using the \cite{2016A&A...588A..25M} post-AGB tracks for an estimated $T_{\rm eff}$= 50-70 kK, \cite{2017ApJ...836...93J} derived a GJJC~1 initial and final mass of 1~\msun~and 0.54-0.59~\msun, respectively. Since the status of this object is still debated and we use updated CS and cluster parameters, we kept this pair in our analysis to try to shed light on its true nature. Our adopted CS temperature agrees with the lower $T_{\rm eff}$ limit reported by \cite{2017ApJ...836...93J} and thus, our adopted lower $T_{\rm eff}$ uncertainty is inconsistent with their estimates. Moreover, for a $T_{\rm eff}$= 50 kK, \cite{2017ApJ...836...93J} estimate L= 3300 $L_{\odot}$, which is higher than our adopted value of L= 2100 $L_{\odot}$. This discrepancy may have been influenced by the different extinction values used to estimate these two luminosities. We find initial and final mass values of $0.78^{+0.02}_{-0.04}$~\msun~and $0.50^{+0.06}_{-0.00}$~\msun, respectively, which are slightly lower but consistent within the errors with the values found by \citep{2017ApJ...836...93J}, confirming their findings. The slow evolution of very low-mass post-AGB stars ($M_{\rm init}$ $<$ 1~\msun) may prevent them from being ionized before the expelled material dissipates, rendering the formation of a visible PN unlikely \cite{2017ApJ...836...93J}. Binary evolution could explain the PNe formed by such low mass stars, though GJJC~1 presents some additional peculiar characteristics, such as a high dust-to-gas ratio and an unusual shape; some of these can also be explained by the presence of a binary companion \cite{2017ApJ...836...93J}. Unfortunately, the CS effective temperature cannot be precisely determined due to the large interstellar extinction uncertainties, though it is estimated to be between $50$-$70$~kK \cite{2017ApJ...836...93J}. Moreover, GJJC~1 is classified as a likely PN \citep{2016JPhCS.728c2008P}, which means that it may not be a PN at all. Deeper PN data and more precise $T_{\rm eff}$ estimates will be required to explore the true nature of this object.

\subsubsection{PN IPHASJ194727.51+241502.1 - OC Teutsch~7 (CS\_ID=2472)}

IPHASJ194727.51+241502.1 (candidate 11 in Figure~\ref{fig:fig1}) is a possible, quasi-stellar (i.e. point-like) PN \citep{2016JPhCS.728c2008P} and its CS is in the group C of \cite{2021A&A...656A..51G}. This is a similar case to PPAJ1756-2311 (see Section 5.1.6). Although the CS initial and final masses are compatible, the quasi-stellar appearance of the PN would imply an extremely compact nebula at the estimated CS distance of 1.19 kpc, calculated from its Gaia parallax (0.84 $\pm$ 0.78). Such a small physical size may not be expected for a CS of 0.62~\msun, which, based on its position among the post-AGB evolutionary tracks (see Figure \ref{fig:fig1}), has already moved past the early post-AGB phase. The large parallax error could imply that the PN's true distance is much larger, explaining its observed small angular size and remaining compatible with the cluster distance of 5.88~kpc. Note that the CS and cluster parallax compatibility takes into account the assigned formal measurement errors (see Section \ref{sec:res}). Additional data and more precise CS distance measurements are needed to assess this candidate CS-cluster pair.

\subsection{Our candidates and the IFMR}\label{sub:ifmr}

The IFMR is vital for stellar evolution studies and provides important information on the chemical enrichment of the ISM. Although it is usually constrained from cluster WDs \citeg{2018ApJ...866...21C}, recent efforts \citeg{2026ApJ...996...90F} aim to improve it using data from the rare confirmed open cluster-PN pairs. Here, we use our best candidate CS-cluster pairs to explore their consistency with previously published IFMRs.

\begin{figure*}[]
    \centering
    \includegraphics[width=0.48\textwidth]{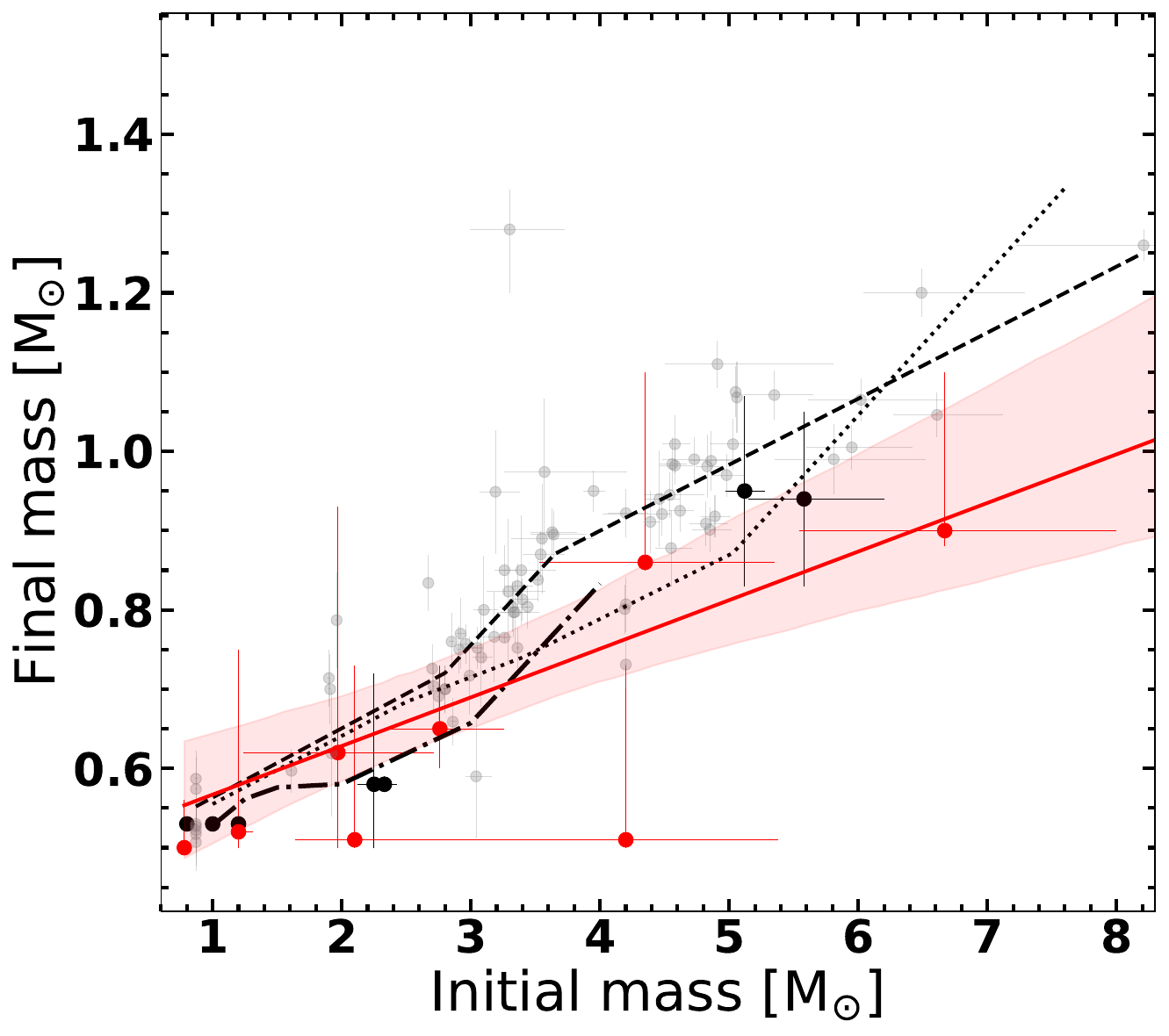}
    \includegraphics[width=0.48\textwidth]{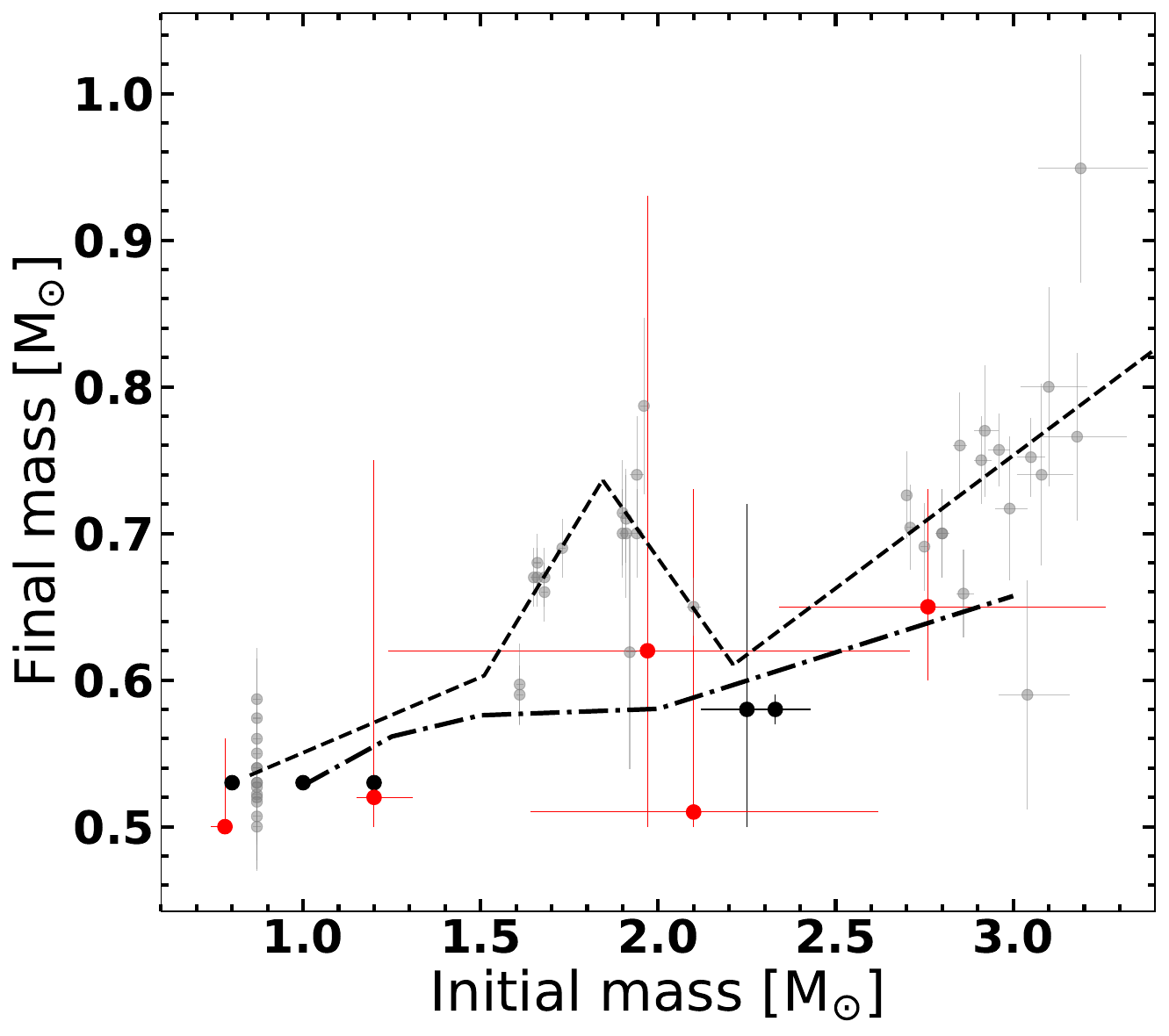}
    \caption{Left panel: Our eight best candidate CS-cluster pairs (red filled circles) plotted along the \cite{2018ApJ...866...21C} semi-empirical IFMR PARSEC fit (black dashed line) and its corresponding points (light grey filled circles) and the \cite{2024MNRAS.527.3602C} fit from stellar population synthesis modeling (black dotted line), together with confirmed or reported cluster-PNe points (black filled circles). The red line shows the best-fit relation determined from the linear regression of our CS initial and final masses, while the surrounding shaded region represents the $1\sigma$ confidence interval derived from 10,000 bootstrap realizations including asymmetric uncertainties in both initial and final masses. Right panel: Same as left, but showing the \cite{2020NatAs...4.1102M} IFMR trend (black dashed line) and its corresponding points (light grey filled circles) that are defined only over the low initial-mass range. In both panels, the dashed-dotted line traces the solar-metallicity initial and final mass points from \cite{2016A&A...588A..25M}. The CS mass values for the confirmed and reported cluster-PNe have been obtained from \cite{2017ApJ...836...93J}, \cite{2019MNRAS.484.3078F}, \cite{2019NatAs...3..851F}, \cite{2025A&A...696A.146F} and \cite{2026ApJ...996...90F}.}
    \label{fig:fig2}
\end{figure*}

In Figure \ref{fig:fig2} we overplot the initial and final masses of our eight best candidate CS-cluster pairs (including the CS of GJJC~1) along the IFMR estimates of \cite{2018ApJ...866...21C} and \cite{2024MNRAS.527.3602C}, presented in the left panel, and \cite{2020NatAs...4.1102M}, presented in the right panel. We also include the solar-metallicity initial and final mass trend from \cite{2016A&A...588A..25M} and the associated points from the remaining confirmed OC-PN pairs \citeg{2026ApJ...996...90F} and reported GC-PN pairs \citep{2017ApJ...836...93J} (see Section \ref{sec:intro}). Moreover, we performed a linear regression of our CS initial and final mass measurements, taking into account their asymmetric uncertainties. The uncertainties of the fitted parameters were derived by a Monte Carlo bootstrap analysis, for which we generated 10,000 realizations of the data by randomly varying each measurement according to its asymmetric uncertainty and performed a linear fit for each realization. The uncertainties were estimated from the 16th and 84th percentiles of the resulting parameter distributions. The computed relation is:
\begin{equation}
M_{\rm fin}
=
(0.506^{+0.092}_{-0.085})
+
(0.061^{+0.029}_{-0.022})M_{\rm init}.
\end{equation}
We also computed the Pearson and Spearman correlation coefficients, obtaining:
\[
r_{\rm Pearson}=0.79,\qquad p=0.019,
\]
and
\[
\rho_{\rm Spearman}=0.72,\qquad p=0.045,
\]
indicating a positive correlation between the initial and final masses, as expected. However, the small size of our sample and the candidate nature of our objects do not allow for an independent determination of the IFMR. The fitted parameters of our best-fit relation are consistent within the uncertainties with those of the IFMR of \cite{2024MNRAS.527.3602C} up to initial masses of 5.03~\msun~and of the semi-empirical IFMR PARSEC fit of \cite{2018ApJ...866...21C} for initial masses below 2.8~\msun~and above 3.65~\msun.

We find that the points corresponding to the majority of our candidates agree within the errors with the previously published IFMR estimates and fall slightly below but parallel to the \cite{2018ApJ...866...21C} and \cite{2020NatAs...4.1102M} general trends, similarly to the IFMR points of the confirmed and reported cluster-PNe. This confirms the systematic effect in CS final masses noted by \cite{2026ApJ...996...90F}, which will be further explored in subsequent studies (Q. A. Parker et al. in preparation). One notable exception is the CS of PHRJ1632-4549, whose estimated final mass is too low for an intermediate-mass progenitor. This could be explained by the considerably large assigned initial and final mass errors, resulting from the propagated uncertainties of the parent catalogues and the propagated $T_{eff}$ uncertainty, which encompasses nearly the full possible $T_{eff}$ range due to inadequate data for estimating a more precise $T_{eff}$ (see Section ~\ref{sec:ext}). For the same reasons the masses of most of our candidates have relatively large errors compared to previously published IFMR data. Our candidate pairs do not contradict, and further support, the "kink" (i.e., a discontinuity in the IFMR slope) found by \cite{2020NatAs...4.1102M} for initial masses between $\sim$ 1.50 and 2.20~\msun. Confirmation of any additional IFMR data points is crucial for stellar evolution studies and thus understanding the chemical enrichment of our Galaxy, since a well-constrained IFMR in this initial-mass range traces the mass loss of the Galaxy's most abundant stars more massive than the Sun. 

\section{Conclusions} \label{sec:conc}

PNe that are physically associated with star clusters are scarce, though important for stellar evolution studies. In this work, we used CS and cluster published catalogues along with Gaia data in an attempt to uncover more of such rare occurrences. We initially uncovered 38 CS-cluster pairs whose kinematic and positional data cannot exclude a possible association (see Table \ref{tab:tab1}). There are 46 additional pairs whose components' angular proximity suggests a possible association but which lack kinematic CS Gaia data. Since these associations cannot be properly explored with the available data, they are presented separately in Table \ref*{tab:tabC1} of the Appendix. Our kinematically consistent candidate pairs were then further explored in terms of interstellar extinction compatibility and consistency with evolutionary models and predictions (see Table \ref{tab:tab3}). Based on the available data and our complete analysis, we find that seven pairs in our sample pass all of our kinematic, extinction, and evolutionary selection criteria, and thus have a higher likelihood of being true associations. Two of these associations are already confirmed or reported in the literature. We explore these seven associations separately, along with the questionable GJJC~1 - NGC~6656 pair, finding that the PN PHRJ1855-0116 - OC UBC~1049 pair is our best candidate for follow-up. The initial and final masses of our best candidates generally agree with the latest IFMR estimates, as well as the points corresponding to confirmed or reported cluster-PNe (see Figure \ref{fig:fig2}).

The source catalogues, which serve as the basis of this work, are not complete \citeg{2023A&A...673A.114H}, and thus more possible pairs could remain hidden from the current study. More data are needed to explore the reported candidate CS-cluster associations. Some of the clusters and CSs in our sample are themselves candidates, so they must first be confirmed. Improved precision in the estimation of cluster, PN and CS kinematic and physical parameters is also required in many cases. In particular, the lack of adequate data for precise CS effective temperature estimates is a major limitation in assessing our candidates based on their evolutionary profiles. Deep spectroscopic PN data and photoionization modeling could allow a more precise estimation of the CS physical properties. Moreover, the nebular internal extinction, which may affect mostly compact PNe \citep{2025MNRAS.543.3035C}, has not been adequately addressed due to insufficient data. Most importantly, for confirming a cluster-PN association, precise radial velocities are needed \citesee{2000ASPC..198..517M} for both the cluster and the PN and/or CS, which requires high-resolution PN and/or CS spectral data \citeg{2026ApJ...996...90F} for stars that are beyond the reach of Gaia. 

Nevertheless, our goal here was to identify the best candidate pairs for follow-up based on the source catalogues and the available data, and not to confirm cluster membership or to claim that pairs absent from our list cannot be truly associated. Further studies are required for a more in-depth analysis of our results.

\begin{acknowledgments}
VF acknowledges support from the Funda\c{c}\~ao de Amparo \`a Pesquisa do Estado do Rio de Janeiro (FAPERJ) through the NOTA 10 postdoctoral program (E-26/200.181/2025, E-26/200.182/2025). LLN thanks Conselho Nacional de Desenvolvimento Científico e Tecnológico (CNPq) for granting the postdoctoral research fellowship 151798/2025-7. DM acknowledges support by the BASAL Center for Astrophysics and Associated Technologies (CATA) through ANID grants ACE210002 and AFB 210003, and by Project Fondecyt Regular 1220724. AC acknowledges the grant 316868/2023-0 of CNPq. DRG acknowledges the grants of the Brazilian agencies FAPERJ (E-26/200.467/2026) and CNPq (315307/2023-4).

This research has made use of NASA's Astrophysics Data System. 
This research has made use of the SIMBAD database, operated at CDS, Strasbourg, France. 
This research has made use of the VizieR catalogue access tool, CDS, Strasbourg, France (doi:10.26093/cds/vizier). 
The original description of the VizieR service was published in \cite{2000A&AS..143...23O}.

This work made use of the University of Hong Kong/Australian Astronomical Observatory/Strasbourg Observatory H-alpha Planetary Nebula (HASH PN) database, hosted by the Laboratory for Space Research at the University of Hong Kong; it is available at http://www.hashpn.space.

This work has made use of data from the European Space Agency (ESA) mission Gaia (https://www.cosmos.esa.int/gaia), processed by the Gaia Data Processing and Analysis Consortium (DPAC, https://www.cosmos.esa.int/web/gaia/dpac/consortium). Funding for the DPAC has been provided by national institutions, in particular the institutions participating in the Gaia Multilateral Agreement.

\end{acknowledgments}

\begin{contribution}

All authors contributed equally.

\end{contribution}

\facilities{Gaia}

\software{TOPCAT \citep{2005ASPC..347...29T},
          PyNeb \citep{2015A&A...573A..42L},
          NumPy \citep{2020Natur.585..357H},
          Matplotlib \citep{2007CSE.....9...90H}}

\clearpage
\appendix
\setcounter{table}{0}
\renewcommand{\thetable}{A\arabic{table}}

\section{Candidate CS-association pairs}

Here, we present the canditate CS-association pairs identified during the initial cross-match after removing pairs with inconsistent radial velocities and parallaxes. Although these pairs were excluded from the subsequent analysis because their potential hosts are classified as stellar associations rather than bound star clusters in the parent catalogues, they are included here as a reference for future studies.

\begin{table}[ht!]
\centering
\setlength{\tabcolsep}{2pt}
\caption{Identified candidate CS-association pairs.}
\label{tab:tabA1}
\begin{tabular}{lccccccccc}
\hline
CS\_ID & PN & cs\_RAJ2000 & cs\_DEJ2000 & Association &
cl\_RAJ2000 & cl\_DEJ2000 & cl\_$R_t$($'$)\tablenotemark{1} &
cl\_orig.\tablenotemark{2} & Separ. ($'$) \\
\hline
225  & BMPJ1820-2441              & 275.026 & -24.690 & HSC\_109    & 274.964 & -24.663 & 10.66  & HR    & 3.75 \\
547  & PHRJ1928+0253              & 292.214 &   2.898 & HSC\_357    & 292.292 &   2.816 & 10.96  & HR    & 6.81 \\
717  & Kn16                       & 305.048 &  24.077 & UPK\_85     & 304.933 &  24.030 & 57.48  & HR    & 6.92 \\
749  & M1-75                      & 301.184 &  31.4570 & Theia\_73   & 301.317 &  31.524 & 129.25 & HR    & 7.90 \\
991  & Te2                        &  85.187 &  31.742 & HSC\_1385   &  85.172 &  31.799 & 62.02  & HR    & 3.49 \\
1024 & Kn40                       &  90.197 &   9.478 & Theia\_3250 &  90.285 &   9.348 & 35.85  & HR    & 9.35 \\
1134 & BMPJ0733-3108              & 113.350 & -31.135 & HSC\_1974   & 113.386 & -31.071 & 76.58  & HR    & 4.21 \\
1232 & VBRC3                      & 145.217 & -56.966 & HSC\_2272   & 145.387 & -56.947 & 8.39   & HR,He & 5.68 \\
1339 & PHRJ1214-6454              & 183.750 & -64.911 & Theia\_5228 & 183.766 & -64.903 & 61.17  & HR    & 0.62 \\
1863 & PHRJ1754-3708              & 268.694 & -37.147 & HSC\_2925   & 268.637 & -37.201 & 11.86  & HR    & 4.20 \\
2143 & TerzN2337                  & 267.188 & -26.722 & HSC\_27     & 267.273 & -26.630 & 17.01  & HR    & 7.15 \\
2347 & M1-54                      & 279.035 & -16.999 & HSC\_189    & 279.065 & -16.996 & 172.27 & HR    & 1.77 \\
2411 & HaTr13                     & 287.010 &   2.357 & HSC\_343    & 286.957 &   2.331 & 110.87 & HR    & 3.52 \\
2475 & IPHASXJ194940.9+261521     & 297.420 &  26.256 & HSC\_507    & 297.452 &  26.201 & 193.70 & HR    & 3.69 \\
2689 & MPAJ1654-4041              & 253.680 & -40.696 & ESO\_332-08 & 253.695 & -40.715 & 10.80  & K     & 1.32 \\
\hline

\end{tabular}
\footnotesize

\tablecomments{\scriptsize
(1) Cluster tidal radius;
(2) Source catalogue for cluster data. HR-\cite{2023A&A...673A.114H}, K–\cite{2013A&A...558A..53K}, He-\cite{2023ApJS..264....8H}. When a cluster is identified in more than one catalogue, employed data are prioritized in the indicated order.
}
\end{table}

\setcounter{table}{0}
\renewcommand{\thetable}{B\arabic{table}}

\section{CS-young cluster matches}

Below are presented the CS-cluster matches that, although they have passed our criteria, the clusters are not expected to host PNe evolved through single star evolution due to their young age ($<50$~Myr). Neveretheless, alternative evolutionary paths, such as binary evolution, may provide a faster evolutionary track.

\begin{table*}
\centering
\scriptsize
\setlength{\tabcolsep}{2pt}
\caption{Identified CS–young cluster matches}
\label{tab:tabB1}

\begin{tabular}{lcccccccccccc}
\hline
CS\_ID & PN & cs\_RAJ2000 & cs\_DEJ2000 & cs\_gr.\tablenotemark{1} & Cluster &
cl\_RAJ2000 & cl\_DEJ2000 & cl\_$R_t$ ($'$)\tablenotemark{2} & cl\_origin\tablenotemark{3} &
separ.($'$) & KF\tablenotemark{4} & EF\tablenotemark{5} \\
\hline

\multicolumn{13}{l}{\textit{CS with kinematic data}}\\
\hline

1042 & IPHASJ063750.80+003644.0 & 99.462 & 0.612 & B & HSC\_1665 & 99.562 & 0.735 & 12.47 & HR & 9.49 & S & EI \\
1231 & IRAS09517-5438           & 148.363 & -54.878 & B & MF\_1 & 148.269 & -55.028 & 12.61 & HR & 9.56 & S & EI \\
1490 & Hen2-114                 & 226.036 & -60.889 & A & UBC\_664 & 226.116 & -60.850 & 13.63 & HR & 3.31 & S & EI \\
1767 & MGE347.2495-00.3759      & 258.222 & -39.773 & A & Teutsch\_85 & 258.309 & -39.709 & 7.86 & HR & 5.58 & S & EI \\
2161 & FrBc2                    & 269.399 & -27.592 & C & UFMG\_89 & 269.279 & -27.607 & 6.45 & HR & 6.42 & S & EC \\
2310 & PHRJ1811-2100            & 272.913 & -21.012 & C & CWNU\_1841 & 272.828 & -20.884 & 10.23 & HR & 9.07 & S & EI \\

\hline

\multicolumn{13}{l}{\textit{CS without kinematic data}}\\
\hline

76   & MGE002.2728-00.9131              & 268.620 & -27.450 & A & HSC\_26      & 268.620 & -27.390 & 14.97 & HR    & 4.19 & & \\
202  & M1-31                            & 268.170 & -22.370 & A & NGC\_6469    & 268.270 & -22.300 & 13.97 & HR,K  & 6.42 & & \\
244  & M1-40                            & 272.110 & -22.280 & B & ASCC\_93     & 272.040 & -22.280 & 9.90  & K     & 3.79 & & \\
273  & PHRJ1806-1956                    & 271.730 & -19.940 & B & FSR\_0039    & 271.760 & -19.910 & 8.40  & K     & 2.47 & & \\
374  & DeGaPe28                         & 274.580 & -12.250 & A & NGC\_6604    & 274.530 & -12.250 & 5.76  & HR,K  & 3.35 & & \\
682  & 2MASSJ19453289+2328105           & 296.390 & 23.470  & A & CWNU\_2167   & 296.340 & 23.440  & -     & He    & 3.53 & & \\
1183 & Hen2-13                          & 130.870 & -46.110 & A & OC\_0483     & 130.860 & -46.150 & 7.35  & HR    & 2.13 & & \\
1355 & He2-86                           & 187.630 & -64.870 & A & NGC\_4463    & 187.460 & -64.800 & 39.11 & HR,K  & 5.97 & & \\
1530 & PHRJ1534-5829                    & 233.690 & -58.500 & A & CWNU\_2248   & 296.620 & 19.360  & -     & He    & 2.48 & & \\
1581 & MPAJ1606-5407                    & 241.660 & -54.120 & A & Theia\_3333  & 241.820 & -54.200 & 11.69 & HR    & 7.13 & & \\
1775 & RPZM8                            & 258.970 & -39.550 & B & NGC\_6318    & 259.050 & -39.430 & 13.70 & HR    & 8.42 & & \\
1881 & PHRJ1735-3333                    & 263.760 & -33.560 & B & Bica\_631    & 263.680 & -33.480 & 23.11 & HR    & 6.21 & & \\
2054 & TerzN41                          & 260.090 & -24.860 & C & OCSN\_1      & 260.160 & -24.860 & 488.40& Q     & 3.78 & & \\
2104 & ShWi2-1                          & 270.610 & -29.420 & C & UFMG\_95     & 270.690 & -29.380 & 16.12 & HR    & 4.56 & & \\
2324 & [UHP2009]VLAG012.1175+01.1965    & 271.920 & -17.860 & C & CWNU\_2480   & 271.920 & -17.890 & 13.97 & HR,He & 1.80 & & \\
2463 & K3-42                            & 294.900 & 20.320  & C & HSC\_457     & 294.820 & 20.170  & 13.41 & HR    & 9.81 & & \\
2488 & K3-52                            & 300.800 & 30.540  & C & HSC\_543     & 300.830 & 30.420  & 14.34 & HR    & 7.40 & & \\
2500 & Pa6                              & 302.420 & 41.240  & C & HSC\_609     & 302.340 & 41.100  & 10.89 & HR    & 9.37 & & \\
2517 & Bl2-1                            & 335.070 & 58.240  & C & NGC\_7261    & 335.050 & 58.130  & 13.48 & HR,K  & 6.73 & & \\
2640 & PHRJ1519-6032                    & 229.850 & -60.540 & C & ASCC\_79     & 229.670 & -60.630 & 30.60 & K     & 7.60 & & \\
2675 & PHRJ1629-4505                    & 247.270 & -45.100 & C & HSC\_2802    & 247.410 & -45.110 & 38.43 & HR    & 5.74 & & \\
2680 & MPAJ1631-4354                    & 247.780 & -43.910 & C & HSC\_2818    & 247.750 & -43.920 & 7.41  & HR    & 1.58 & & \\
2733 & MGE354.0020-01.8419              & 264.520 & -35.000 & C & NGC\_6396    & 264.400 & -35.020 & 11.53 & HR,K  & 5.99 & & \\
2750 & PHRJ1718-3055                    & 259.690 & -30.920 & C & CWNU\_170    & 259.630 & -30.950 & 7.01  & HR    & 3.57 & & \\
2835 & MGE358.8655+01.2862              & 264.470 & -29.220 & C & HSC\_2980    & 264.510 & -29.120 & 16.59 & HR    & 6.31 & & \\
2865 & JaSt72                           & 267.620 & -30.060 & C & NGC\_6451    & 267.670 & -30.210 & 19.43 & HR,K  & 9.40 & & \\

\hline
\end{tabular}

\tablecomments{\scriptsize
(1) CS group from \GStwentyone;
(2) Cluster tidal radius;
(3) Source catalogue for cluster data. HR-\HRtwentythree, K–\Kharthirteen, He-\cite{2023ApJS..264....8H}, Q-\cite{2023ApJS..265...12Q}. When a cluster is identified in more than one catalogue, employed data are prioritized in the indicated order; 
(4) Kinematic consistency flag: S- strong associations, M- moderate associations, W- weak associations; 
(5) Extinction consistency flag: EC- Compatible associations based on agreement of cluster and PN interstellar extinctions;
EI- Cluster and PN interstellar extinctions are incompatible;}

\end{table*}

\clearpage

\setcounter{table}{0}
\renewcommand{\thetable}{C\arabic{table}}

\section{Candidate pairs excluded from our sample because of the lack of central star kinematic data.}

In Table C1 below we present all candidate pairs that have been excluded from our sample due to their lack of CS kinematic dat﻿a.

\begin{table}[ht!]
\centering
\scriptsize
\setlength{\tabcolsep}{2pt} 
\caption{Candidate CS-cluster pairs with no CS kinematic data. All clusters in this sample are OCs apart from NGC~6540 (CS\_ID=2187), which is a GC.}
\label{tab:tabC1}
\begin{tabular}{l c c c c c c c c c c}
\hline
CS\_ID & PN\_name & cs\_RAJ2000 & cs\_DEJ2000 & cs\_gr\tablenotemark{1} & cl\_name & cl\_RAJ2000 & cl\_DEJ2000 & cl\_$R_t$($'$)\tablenotemark{2} & cl\_orig.\tablenotemark{3} & separ. (') \\
\hline
51 & JaSt~82 & 268.0220 & -28.0965 & B & UFMG~83 & 268.1100 & -28.1821 & 9.98 & HR & 6.93 \\
73 & M~3-20 & 269.8305 & -28.2301 & A & Trumpler~31 & 269.9320 & -28.2070 & 9.90 & K & 5.54 \\
156 & PHRJ1759-2501 & 269.7698 & -25.0314 & A & HSC~62 & 269.8259 & -25.0763 & 12.99 & HR & 4.07 \\
160 & Pe~1-9 & 266.4032 & -23.0406 & B & HSC~65 & 266.3506 & -23.0948 & 11.05 & HR & 4.36 \\
250 & MGE008.8168+02.2819 & 269.2043 & -20.1933 & B & Theia~1854 & 269.2384 & -20.3142 & 19.70 & HR & 7.51 \\
307 & FBP~2 & 273.3003 & -18.3355 & B & UPK~5 & 273.2102 & -18.3261 & 211.27 & HR & 5.16 \\
316 & Sh~2-42 & 272.5584 & -16.7946 & A & Gulliver~15 & 272.5981 & -16.7226 & 13.32 & HR,K & 4.88 \\
328 & MSX6CG014.2361+00.2138 & 273.8935 & -16.4745 & A & Theia~1756 & 274.0201 & -16.4104 & 46.78 & HR & 8.24 \\
358 & MPAJ1828-1516 & 277.0108 & -15.2732 & A & Theia~2510 & 277.0395 & -15.3314 & 25.78 & HR & 3.87 \\
452 & PHRJ1843-0541 & 280.7948 & -5.6984 & B & HSC~266 & 280.7894 & -5.7876 & 6.65 & HR & 5.36 \\
475 & MPAJ1836-0227 & 279.0446 & -2.4539 & A & HSC~290 & 278.9659 & -2.3345 & 15.81 & HR & 8.58 \\
635 & IPHASXJ193009.3+192129 & 292.5400 & 19.3562 & A & UBC~1063 & 292.4175 & 19.4418 & 9.52 & HR & 8.63 \\
655 & IPHASXJ194533.6+210808 & 296.3896 & 21.1358 & A & LP~321 & 296.3863 & 21.1628 & 20.52 & HR & 1.63 \\
1097 & PHRJ0724-1757 & 111.1805 & -17.9614 & A & HSC~1838 & 111.0144 & -17.9450 & 10.24 & HR & 9.53 \\
1196 & PM~1-47 & 138.3876 & -52.0449 & B & UBC~1458 & 138.3721 & -51.9052 & 16.04 & HR & 8.40 \\
1419 & MaC~1-2 & 208.6141 & -64.9923 & A & Loden~995 & 208.9408 & -64.9486 & 20.74 & HR,K & 8.70 \\
1502 & PrePa~3 & 227.2351 & -59.5545 & A & CWNU~1275 & 227.2161 & -59.6453 & 9.29 & HR & 5.48 \\
1510 & Pa~65 & 226.6365 & -56.4347 & B & FSR~1697 & 226.6750 & -56.5170 & 6.30 & K & 5.10 \\
1569 & He~2-143 & 240.2465 & -55.0944 & A & CWNU~2570 & 240.1363 & -55.0607 & 4.31 & HR & 4.29 \\
1742 & Vd~1-4 & 252.6056 & -39.1386 & A & FSR~1746 & 252.6750 & -39.0800 & 8.40 & K & 4.78 \\
1770 & PPAJ1704-3824 & 256.2464 & -38.4012 & A & ESO~332-20 & 256.2280 & -38.3170 & 6.60 & K & 5.13 \\
1774 & Vd~1-8 & 256.1408 & -37.8875 & A & NGC~6281 & 256.1888 & -37.9407 & 55.77 & HR,K,Q & 3.92 \\
1938 & PPAJ1734-3102 & 263.6432 & -31.0355 & A & HSC~2959 & 263.5532 & -31.1180 & 8.97 & HR & 6.77 \\
2014 & Hb~5 & 266.9840 & -29.9942 & A & Ruprecht~130 & 266.8948 & -30.0979 & 9.31 & HR & 7.76 \\
2060 & M~2-20 & 268.6057 & -29.6024 & C & HSC~2 & 268.6274 & -29.5311 & 16.79 & HR & 4.43 \\
2104 & ShWi~2-1 & 270.6102 & -29.4182 & C & UFMG~95 & 270.6882 & -29.3841 & 16.12 & HR & 4.56 \\
2131 & [GKF2010]MN60 & 265.5313 & -26.0369 & C & HSC~20 & 265.5656 & -26.1113 & 6.83 & HR & 4.83 \\
2156 & PHRJ1801-2809 & 270.4383 & -28.1603 & C & CWNU~1494 & 270.5164 & -28.2890 & 14.15 & HR & 8.76 \\
2187 & PHRJ1806-2747 & 271.6160 & -27.7878 & C & NGC~6540 & 271.5358 & -27.7653 & 4.75 & Har\tablenotemark{4} & 4.47 \\
2329 & Pa~84 & 269.5841 & -16.2061 & C & HSC~156 & 269.6887 & -16.2041 & 15.39 & HR & 6.03 \\
2441 & K~4-16 & 286.2142 & 15.7935 & C & UPK~54 & 286.2544 & 15.7219 & 28.06 & HR & 4.89 \\
2450 & IPHASXJ191716.5+181518 & 289.3200 & 18.2570 & C & Theia~836 & 289.3033 & 18.4197 & 79.09 & HR & 9.81 \\
2460 & Kn~43 & 295.2815 & 19.1420 & C & UBC~124 & 295.4163 & 19.0406 & 13.13 & HR & 9.77 \\
2558 & PHRJ0724-2021 & 111.0573 & -20.3655 & C & PHOC~16 & 111.0504 & -20.2084 & 14.11 & HR & 9.43 \\
2564 & M~3-5 & 120.6208 & -27.6983 & C & UBC~640 & 120.6237 & -27.8300 & 10.07 & HR & 7.91 \\
2579 & Hen~2-29 & 141.1910 & -54.6043 & C & BH~66 & 141.3167 & -54.7194 & 9.64 & HR & 8.17 \\
2617 & MPAJ1326-6407 & 201.6361 & -64.1197 & C & Teutsch~110 & 201.6366 & -64.1057 & 3.20 & HR & 0.84 \\
2620 & PHRJ1326-6103 & 201.5872 & -61.0527 & C & CWNU~1442 & 201.6773 & -60.9197 & 8.43 & HR & 8.40 \\
2636 & PHRJ1457-5812 & 224.3957 & -58.2035 & C & CWNU~1528 & 224.4900 & -58.0551 & 12.64 & HR & 9.39 \\
2645 & Ro~2 & 233.8583 & -57.6757 & C & UFMG~34 & 233.8402 & -57.6560 & 15.05 & HR & 1.32 \\
2664 & IRAS16097-5257 & 243.4235 & -53.0809 & C & Theia~3669 & 243.4974 & -52.9652 & 16.13 & HR & 7.44 \\
2685 & H~1-3 & 253.3809 & -42.6574 & C & Theia~659 & 253.3758 & -42.5887 & 81.79 & HR & 4.13 \\
2688 & IRAS16515-4050 & 253.7519 & -40.9263 & C & BH~202 & 253.7872 & -40.9567 & 18.21 & HR,K & 2.43 \\
2704 & PreMo~2 & 260.1516 & -38.2138 & C & HSC~2885 & 260.2207 & -38.2254 & 17.38 & HR & 3.33 \\
2725 & RPZM~23 & 263.3444 & -35.1845 & C & UFMG~87 & 263.2987 & -35.1361 & 5.55 & HR & 3.67 \\
2748 & MPAJ1744-3444 & 266.1814 & -34.7407 & C & Ruprecht~128 & 266.0663 & -34.8767 & 15.92 & HR & 9.94 \\
\hline
\end{tabular}
\footnotesize

\tablecomments{\scriptsize
(1) CS group from \cite{2021A&A...656A..51G};
(2) Cluster tidal radius;
(3) Source catalogue for cluster data. HR-\cite{2023A&A...673A.114H}, K–\cite{2013A&A...558A..53K}, Q-\cite{2023ApJS..265...12Q}, Har-\cite{1996AJ....112.1487H}. When a cluster is identified in more than one catalogue, employed data are prioritized in the indicated order; 
(4) [Fe/H] from \cite{1996AJ....112.1487H}, cluster radius from \cite{2011CQGra..28h5016W}, kinematic data from \cite{2021MNRAS.505.5978V}.
}

\end{table}

\bibliography{sample701}{}

@ARTICLE{2026ApJ...996...90F,
       author = {{Fragkou}, Vasiliki and {Parker}, Quentin A. and {Gon{\c{c}}alves}, Denise R.},
        title = "{PHR J1724-3859: A Bipolar Planetary Nebula in Open Cluster Trumpler 25}",
      journal = {\apj},
         year = 2026,
        month = jan,
       volume = {996},
       number = {1},
          eid = {90},
        pages = {90},
          doi = {10.3847/1538-4357/ae1f0c},
archivePrefix = {arXiv},
       eprint = {2511.15115},
 primaryClass = {astro-ph.SR},
       adsurl = {https://ui.adsabs.harvard.edu/abs/2026ApJ...996...90F}
}

@ARTICLE{2025MNRAS.543.3035C,
       author = {{Csukai}, Alexander and {Zijlstra}, Albert A. and {McDonald}, Iain and {De Marco}, Orsola},
        title = "{Central-star extinctions towards planetary nebulae}",
      journal = {\mnras},
         year = 2025,
        month = nov,
       volume = {543},
       number = {3},
        pages = {3035-3054},
          doi = {10.1093/mnras/staf1552},
archivePrefix = {arXiv},
       eprint = {2509.10621},
 primaryClass = {astro-ph.SR},
       adsurl = {https://ui.adsabs.harvard.edu/abs/2025MNRAS.543.3035C}
}

@ARTICLE{2025A&A...702A..79M,
       author = {{Minniti}, Dante and {Fragkou}, Vasiliki and {Alonso-Garc{\'\i}a}, Javier and {Majaess}, Daniel and {Cortesi}, Arianna},
        title = "{Near-infrared photometry of the central stars of planetary nebulae with the VVVX survey}",
      journal = {\aap},
         year = 2025,
        month = oct,
       volume = {702},
          eid = {A79},
        pages = {A79},
          doi = {10.1051/0004-6361/202555744},
archivePrefix = {arXiv},
       eprint = {2510.02179},
 primaryClass = {astro-ph.SR},
       adsurl = {https://ui.adsabs.harvard.edu/abs/2025A&A...702A..79M}
}

@ARTICLE{2025A&A...696A.146F,
       author = {{Fragkou}, Vasiliki and {V{\'a}zquez}, Roberto and {Parker}, Quentin A. and {Gon{\c{c}}alves}, Denise R. and {Lomel{\'\i}-N{\'u}{\~n}ez}, Luis},
        title = "{The physical association of planetary nebula NGC 2818 with open cluster NGC 2818A}",
      journal = {\aap},
         year = 2025,
        month = apr,
       volume = {696},
          eid = {A146},
        pages = {A146},
          doi = {10.1051/0004-6361/202453031},
archivePrefix = {arXiv},
       eprint = {2503.19287},
 primaryClass = {astro-ph.SR},
       adsurl = {https://ui.adsabs.harvard.edu/abs/2025A&A...696A.146F}
}

@ARTICLE{2024A&A...688A...1C,
       author = {{Castro-Ginard}, Alfred and {Penoyre}, Zephyr and {Casey}, Andrew R. and {Brown}, Anthony G.~A. and {Belokurov}, Vasily and {Cantat-Gaudin}, Tristan and {Drimmel}, Ronald and {Fouesneau}, Morgan and {Khanna}, Shourya and {Kurbatov}, Evgeny P. and {Price-Whelan}, Adrian M. and {Rix}, Hans-Walter and {Smart}, Richard L.},
        title = "{Gaia DR3 detectability of unresolved binary systems}",
      journal = {\aap},
         year = 2024,
        month = aug,
       volume = {688},
          eid = {A1},
        pages = {A1},
          doi = {10.1051/0004-6361/202450172},
archivePrefix = {arXiv},
       eprint = {2404.14127},
 primaryClass = {astro-ph.GA},
       adsurl = {https://ui.adsabs.harvard.edu/abs/2024A&A...688A...1C}
}

@ARTICLE{2024A&A...686A..42H,
       author = {{Hunt}, Emily L. and {Reffert}, Sabine},
        title = "{Improving the open cluster census. III. Using cluster masses, radii, and dynamics to create a cleaned open cluster catalogue}",
      journal = {\aap},
         year = 2024,
        month = jun,
       volume = {686},
          eid = {A42},
        pages = {A42},
          doi = {10.1051/0004-6361/202348662},
archivePrefix = {arXiv},
       eprint = {2403.05143},
 primaryClass = {astro-ph.GA},
       adsurl = {https://ui.adsabs.harvard.edu/abs/2024A&A...686A..42H}
}

@ARTICLE{2024MNRAS.527.3602C,
       author = {{Cunningham}, Tim and {Tremblay}, Pier-Emmanuel and {W. O'Brien}, Mairi},
        title = "{Initial-final mass relation from white dwarfs within 40 pc}",
      journal = {\mnras},
         year = 2024,
        month = jan,
       volume = {527},
       number = {2},
        pages = {3602-3611},
          doi = {10.1093/mnras/stad3275},
archivePrefix = {arXiv},
       eprint = {2310.15410},
 primaryClass = {astro-ph.SR},
       adsurl = {https://ui.adsabs.harvard.edu/abs/2024MNRAS.527.3602C}
}

@ARTICLE{2023A&A...673A.114H,
       author = {{Hunt}, Emily L. and {Reffert}, Sabine},
        title = "{Improving the open cluster census. II. An all-sky cluster catalogue with Gaia DR3}",
      journal = {\aap},
         year = 2023,
        month = may,
       volume = {673},
          eid = {A114},
        pages = {A114},
          doi = {10.1051/0004-6361/202346285},
archivePrefix = {arXiv},
       eprint = {2303.13424},
 primaryClass = {astro-ph.GA},
       adsurl = {https://ui.adsabs.harvard.edu/abs/2023A&A...673A.114H}
}

@ARTICLE{2023MNRAS.520..773R,
       author = {{Ritter}, Andreas and {Parker}, Q.~A. and {Sabin}, L. and {Le D{\^u}}, P. and {Mulato}, L. and {Patchick}, D.},
        title = "{Grantecan spectroscopic observations and confirmations of planetary nebulae candidates in the Northern Galactic Plane}",
      journal = {\mnras},
         year = 2023,
        month = mar,
       volume = {520},
       number = {1},
        pages = {773-781},
          doi = {10.1093/mnras/stac2896},
archivePrefix = {arXiv},
       eprint = {2210.07581},
 primaryClass = {astro-ph.SR},
       adsurl = {https://ui.adsabs.harvard.edu/abs/2023MNRAS.520..773R}
}

@ARTICLE{2023ApJS..265...12Q,
       author = {{Qin}, Songmei and {Zhong}, Jing and {Tang}, Tong and {Chen}, Li},
        title = "{Hunting for Neighboring Open Clusters with Gaia DR3: 101 New Open Clusters within 500 pc}",
      journal = {\apjs},
         year = 2023,
        month = mar,
       volume = {265},
       number = {1},
          eid = {12},
        pages = {12},
          doi = {10.3847/1538-4365/acadd6},
archivePrefix = {arXiv},
       eprint = {2212.11034},
 primaryClass = {astro-ph.SR},
       adsurl = {https://ui.adsabs.harvard.edu/abs/2023ApJS..265...12Q}
}

@ARTICLE{2023ApJS..264....8H,
       author = {{He}, Zhihong and {Liu}, Xiaochen and {Luo}, Yangping and {Wang}, Kun and {Jiang}, Qingquan},
        title = "{Unveiling Hidden Stellar Aggregates in the Milky Way: 1656 New Star Clusters Found in Gaia EDR3}",
      journal = {\apjs},
         year = 2023,
        month = jan,
       volume = {264},
       number = {1},
          eid = {8},
        pages = {8},
          doi = {10.3847/1538-4365/ac9af8},
archivePrefix = {arXiv},
       eprint = {2209.08504},
 primaryClass = {astro-ph.GA},
       adsurl = {https://ui.adsabs.harvard.edu/abs/2023ApJS..264....8H}
}

@ARTICLE{2022A&A...667A.148G,
       author = {{Gaia Collaboration} and {Klioner}, S.~A. and {Lindegren}, L. and {Mignard}, F. and {Hern{\'a}ndez}, J. and {Ramos-Lerate}, M. and {Bastian}, U. and {Biermann}, M. and {Bombrun}, A. and {de Torres}, A. and {Gerlach}, E. and {Geyer}, R. and {Hilger}, T. and {Hobbs}, D. and {Lammers}, U.~L. and {McMillan}, P.~J. and {Steidelm{\"u}ller}, H. and {Teyssier}, D. and {Raiteri}, C.~M. and {Bartolom{\'e}}, S. and {Bernet}, M. and {Casta{\~n}eda}, J. and {Clotet}, M. and {Davidson}, M. and {Fabricius}, C. and {Garralda Torres}, N. and {Gonz{\'a}lez-Vidal}, J.~J. and {Portell}, J. and {Rowell}, N. and {Torra}, F. and {Torra}, J. and {Brown}, A.~G.~A. and {Vallenari}, A. and {Prusti}, T. and {de Bruijne}, J.~H.~J. and {Arenou}, F. and {Babusiaux}, C. and {Creevey}, O.~L. and {Ducourant}, C. and {Evans}, D.~W. and {Eyer}, L. and {Guerra}, R. and {Hutton}, A. and {Jordi}, C. and {Luri}, X. and {Panem}, C. and {Pourbaix}, D. and {Randich}, S. and {Sartoretti}, P. and {Soubiran}, C. and {Tanga}, P. and {Walton}, N.~A. and {Bailer-Jones}, C.~A.~L. and {Drimmel}, R. and {Jansen}, F. and {Katz}, D. and {Lattanzi}, M.~G. and {van Leeuwen}, F. and {Bakker}, J. and {Cacciari}, C. and {De Angeli}, F. and {Fouesneau}, M. and {Fr{\'e}mat}, Y. and {Galluccio}, L. and {Guerrier}, A. and {Heiter}, U. and {Masana}, E. and {Messineo}, R. and {Mowlavi}, N. and {Nicolas}, C. and {Nienartowicz}, K. and {Pailler}, F. and {Panuzzo}, P. and {Riclet}, F. and {Roux}, W. and {Seabroke}, G.~M. and {Sordo}, R. and {Th{\'e}venin}, F. and {Gracia-Abril}, G. and {Altmann}, M. and {Andrae}, R. and {Audard}, M. and {Bellas-Velidis}, I. and {Benson}, K. and {Berthier}, J. and {Blomme}, R. and {Burgess}, P.~W. and {Busonero}, D. and {Busso}, G. and {C{\'a}novas}, H. and {Carry}, B. and {Cellino}, A. and {Cheek}, N. and {Clementini}, G. and {Damerdji}, Y. and {de Teodoro}, P. and {Nu{\~n}ez Campos}, M. and {Delchambre}, L. and {Dell'Oro}, A. and {Esquej}, P. and {Fern{\'a}ndez-Hern{\'a}ndez}, J. and {Fraile}, E. and {Garabato}, D. and {Garc{\'\i}a-Lario}, P. and {Gosset}, E. and {Haigron}, R. and {Halbwachs}, J.-L. and {Hambly}, N.~C. and {Harrison}, D.~L. and {Hestroffer}, D. and {Hodgkin}, S.~T. and {Holl}, B. and {Jan{\ss}en}, K. and {Jevardat de Fombelle}, G. and {Jordan}, S. and {Krone-Martins}, A. and {Lanzafame}, A.~C. and {L{\"o}ffler}, W. and {Marchal}, O. and {Marrese}, P.~M. and {Moitinho}, A. and {Muinonen}, K. and {Osborne}, P. and {Pancino}, E. and {Pauwels}, T. and {Recio-Blanco}, A. and {Reyl{\'e}}, C. and {Riello}, M. and {Rimoldini}, L. and {Roegiers}, T. and {Rybizki}, J. and {Sarro}, L.~M. and {Siopis}, C. and {Smith}, M. and {Sozzetti}, A. and {Utrilla}, E. and {van Leeuwen}, M. and {Abbas}, U. and {{\'A}brah{\'a}m}, P. and {Abreu Aramburu}, A. and {Aerts}, C. and {Aguado}, J.~J. and {Ajaj}, M. and {Aldea-Montero}, F. and {Altavilla}, G. and {{\'A}lvarez}, M.~A. and {Alves}, J. and {Anderson}, R.~I. and {Anglada Varela}, E. and {Antoja}, T. and {Baines}, D. and {Baker}, S.~G. and {Balaguer-N{\'u}{\~n}ez}, L. and {Balbinot}, E. and {Balog}, Z. and {Barache}, C. and {Barbato}, D. and {Barros}, M. and {Barstow}, M.~A. and {Bassilana}, J.-L. and {Bauchet}, N. and {Becciani}, U. and {Bellazzini}, M. and {Berihuete}, A. and {Bertone}, S. and {Bianchi}, L. and {Binnenfeld}, A. and {Blanco-Cuaresma}, S. and {Boch}, T. and {Bossini}, D. and {Bouquillon}, S. and {Bragaglia}, A. and {Bramante}, L. and {Breedt}, E. and {Bressan}, A. and {Brouillet}, N. and {Brugaletta}, E. and {Bucciarelli}, B. and {Burlacu}, A. and {Butkevich}, A.~G. and {Buzzi}, R. and {Caffau}, E. and {Cancelliere}, R. and {Cantat-Gaudin}, T. and {Carballo}, R. and {Carlucci}, T. and {Carnerero}, M.~I. and {Carrasco}, J.~M. and {Casamiquela}, L. and {Castellani}, M. and {Castro-Ginard}, A. and {Chaoul}, L. and {Charlot}, P. and {Chemin}, L. and {Chiaramida}, V. and {Chiavassa}, A. and {Chornay}, N. and {Comoretto}, G. and {Contursi}, G. and {Cooper}, W.~J.},
        title = "{Gaia Early Data Release 3. The celestial reference frame (Gaia-CRF3)}",
      journal = {\aap},
         year = 2022,
        month = nov,
       volume = {667},
          eid = {A148},
        pages = {A148},
          doi = {10.1051/0004-6361/202243483},
archivePrefix = {arXiv},
       eprint = {2204.12574},
 primaryClass = {astro-ph.IM},
       adsurl = {https://ui.adsabs.harvard.edu/abs/2022A&A...667A.148G}
}

@ARTICLE{2022Galax..10...44F,
       author = {{Fragkou}, Vasiliki and {Parker}, Quentin A. and {Zijlstra}, Albert A. and {Crause}, Lisa and {Sabin}, Laurence and {V{\'a}zquez}, Roberto},
        title = "{Further Studies of the Association of Planetary Nebula BMP J16135406 with Galactic Open Cluster NGC 6067}",
      journal = {Galaxies},
         year = 2022,
        month = mar,
       volume = {10},
       number = {2},
          eid = {44},
        pages = {44},
          doi = {10.3390/galaxies10020044},
       adsurl = {https://ui.adsabs.harvard.edu/abs/2022Galax..10...44F}
}

@ARTICLE{2022ApJ...935L..35F,
       author = {{Fragkou}, Vasiliki and {Parker}, Quentin A. and {Zijlstra}, Albert A. and {V{\'a}zquez}, Roberto and {Sabin}, Laurence and {Rechy-Garcia}, Jackeline Suzett},
        title = "{The Planetary Nebula in the 500 Myr Old Open Cluster M37}",
      journal = {\apjl},
         year = 2022,
        month = aug,
       volume = {935},
       number = {2},
          eid = {L35},
        pages = {L35},
          doi = {10.3847/2041-8213/ac88c1},
archivePrefix = {arXiv},
       eprint = {2208.06101},
 primaryClass = {astro-ph.SR},
       adsurl = {https://ui.adsabs.harvard.edu/abs/2022ApJ...935L..35F}
}

@ARTICLE{2022Galax..10...32P,
       author = {{Parker}, Quentin A. and {Xiang}, Zou and {Ritter}, Andreas},
        title = "{A Preliminary Investigation of CSPN in the HASH Database}",
      journal = {Galaxies},
         year = 2022,
        month = feb,
       volume = {10},
       number = {1},
          eid = {32},
        pages = {32},
          doi = {10.3390/galaxies10010032},
       adsurl = {https://ui.adsabs.harvard.edu/abs/2022Galax..10...32P}
}

@ARTICLE{2021A&A...656A..51G,
       author = {{Gonz{\'a}lez-Santamar{\'\i}a}, I. and {Manteiga}, M. and {Manchado}, A. and {Ulla}, A. and {Dafonte}, C. and {L{\'o}pez Varela}, P.},
        title = "{Planetary nebulae in Gaia EDR3: Central star identification, properties, and binarity}",
      journal = {\aap},
         year = 2021,
        month = dec,
       volume = {656},
          eid = {A51},
        pages = {A51},
          doi = {10.1051/0004-6361/202141916},
archivePrefix = {arXiv},
       eprint = {2109.12114},
 primaryClass = {astro-ph.GA},
       adsurl = {https://ui.adsabs.harvard.edu/abs/2021A&A...656A..51G}
}

@ARTICLE{2021MNRAS.505.5978V,
       author = {{Vasiliev}, Eugene and {Baumgardt}, Holger},
        title = "{Gaia EDR3 view on galactic globular clusters}",
      journal = {\mnras},
         year = 2021,
        month = aug,
       volume = {505},
       number = {4},
        pages = {5978-6002},
          doi = {10.1093/mnras/stab1475},
archivePrefix = {arXiv},
       eprint = {2102.09568},
 primaryClass = {astro-ph.GA},
       adsurl = {https://ui.adsabs.harvard.edu/abs/2021MNRAS.505.5978V}
}

@ARTICLE{2021MNRAS.504..356D,
       author = {{Dias}, W.~S. and {Monteiro}, H. and {Moitinho}, A. and {L{\'e}pine}, J.~R.~D. and {Carraro}, G. and {Paunzen}, E. and {Alessi}, B. and {Villela}, L.},
        title = "{Updated parameters of 1743 open clusters based on Gaia DR2}",
      journal = {\mnras},
         year = 2021,
        month = jun,
       volume = {504},
       number = {1},
        pages = {356-371},
          doi = {10.1093/mnras/stab770},
archivePrefix = {arXiv},
       eprint = {2103.12829},
 primaryClass = {astro-ph.SR},
       adsurl = {https://ui.adsabs.harvard.edu/abs/2021MNRAS.504..356D}
}

@ARTICLE{2021A&A...649A...5F,
       author = {{Fabricius}, C. and {Luri}, X. and {Arenou}, F. and {Babusiaux}, C. and {Helmi}, A. and {Muraveva}, T. and {Reyl{\'e}}, C. and {Spoto}, F. and {Vallenari}, A. and {Antoja}, T. and {Balbinot}, E. and {Barache}, C. and {Bauchet}, N. and {Bragaglia}, A. and {Busonero}, D. and {Cantat-Gaudin}, T. and {Carrasco}, J.~M. and {Diakit{\'e}}, S. and {Fabrizio}, M. and {Figueras}, F. and {Garcia-Gutierrez}, A. and {Garofalo}, A. and {Jordi}, C. and {Kervella}, P. and {Khanna}, S. and {Leclerc}, N. and {Licata}, E. and {Lambert}, S. and {Marrese}, P.~M. and {Masip}, A. and {Ramos}, P. and {Robichon}, N. and {Robin}, A.~C. and {Romero-G{\'o}mez}, M. and {Rubele}, S. and {Weiler}, M.},
        title = "{Gaia Early Data Release 3. Catalogue validation}",
      journal = {\aap},
         year = 2021,
        month = may,
       volume = {649},
          eid = {A5},
        pages = {A5},
          doi = {10.1051/0004-6361/202039834},
archivePrefix = {arXiv},
       eprint = {2012.06242},
 primaryClass = {astro-ph.GA},
       adsurl = {https://ui.adsabs.harvard.edu/abs/2021A&A...649A...5F}
}

@ARTICLE{2020A&A...644A.173G,
       author = {{Gonz{\'a}lez-Santamar{\'\i}a}, I. and {Manteiga}, M. and {Manchado}, A. and {G{\'o}mez-Mu{\~n}oz}, M.~A. and {Ulla}, A. and {Dafonte}, C.},
        title = "{Wide binaries in planetary nebulae with Gaia DR2}",
      journal = {\aap},
         year = 2020,
        month = dec,
       volume = {644},
          eid = {A173},
        pages = {A173},
          doi = {10.1051/0004-6361/202039422},
archivePrefix = {arXiv},
       eprint = {2011.06357},
 primaryClass = {astro-ph.SR},
       adsurl = {https://ui.adsabs.harvard.edu/abs/2020A&A...644A.173G}
}

@ARTICLE{2020Natur.585..357H,
       author = {{Harris}, Charles R. and {Millman}, K. Jarrod and {van der Walt}, St{\'e}fan J. and {Gommers}, Ralf and {Virtanen}, Pauli and {Cournapeau}, David and {Wieser}, Eric and {Taylor}, Julian and {Berg}, Sebastian and {Smith}, Nathaniel J. and {Kern}, Robert and {Picus}, Matti and {Hoyer}, Stephan and {van Kerkwijk}, Marten H. and {Brett}, Matthew and {Haldane}, Allan and {del R{\'\i}o}, Jaime Fern{\'a}ndez and {Wiebe}, Mark and {Peterson}, Pearu and {G{\'e}rard-Marchant}, Pierre and {Sheppard}, Kevin and {Reddy}, Tyler and {Weckesser}, Warren and {Abbasi}, Hameer and {Gohlke}, Christoph and {Oliphant}, Travis E.},
        title = "{Array programming with NumPy}",
      journal = {\nat},
         year = 2020,
        month = sep,
       volume = {585},
       number = {7825},
        pages = {357-362},
          doi = {10.1038/s41586-020-2649-2},
archivePrefix = {arXiv},
       eprint = {2006.10256},
 primaryClass = {cs.MS},
       adsurl = {https://ui.adsabs.harvard.edu/abs/2020Natur.585..357H}
}

@ARTICLE{2020A&A...640A..10W,
       author = {{Weidmann}, W.~A. and {Mari}, M.~B. and {Schmidt}, E.~O. and {Gaspar}, G. and {Miller Bertolami}, M.~M. and {Oio}, G.~A. and {Guti{\'e}rrez-Soto}, L.~A. and {Volpe}, M.~G. and {Gamen}, R. and {Mast}, D.},
        title = "{Catalogue of the central stars of planetary nebulae. Expanded edition}",
      journal = {\aap},
         year = 2020,
        month = aug,
       volume = {640},
          eid = {A10},
        pages = {A10},
          doi = {10.1051/0004-6361/202037998},
archivePrefix = {arXiv},
       eprint = {2005.10368},
 primaryClass = {astro-ph.GA},
       adsurl = {https://ui.adsabs.harvard.edu/abs/2020A&A...640A..10W}
}

@ARTICLE{2020NatAs...4.1102M,
       author = {{Marigo}, Paola and {Cummings}, Jeffrey D. and {Curtis}, Jason Lee and {Kalirai}, Jason and {Chen}, Yang and {Tremblay}, Pier-Emmanuel and {Ramirez-Ruiz}, Enrico and {Bergeron}, Pierre and {Bladh}, Sara and {Bressan}, Alessandro and {Girardi}, L{\'e}o and {Pastorelli}, Giada and {Trabucchi}, Michele and {Cheng}, Sihao and {Aringer}, Bernhard and {Tio}, Piero Dal},
        title = "{Carbon star formation as seen through the non-monotonic initial-final mass relation}",
      journal = {Nature Astronomy},
         year = 2020,
        month = jul,
       volume = {4},
        pages = {1102-1110},
          doi = {10.1038/s41550-020-1132-1},
archivePrefix = {arXiv},
       eprint = {2007.04163},
 primaryClass = {astro-ph.SR},
       adsurl = {https://ui.adsabs.harvard.edu/abs/2020NatAs...4.1102M}
}

@ARTICLE{2020A&A...638A.103C,
       author = {{Chornay}, N. and {Walton}, N.~A.},
        title = "{Searching for central stars of planetary nebulae in Gaia DR2}",
      journal = {\aap},
         year = 2020,
        month = jun,
       volume = {638},
          eid = {A103},
        pages = {A103},
          doi = {10.1051/0004-6361/202037554},
archivePrefix = {arXiv},
       eprint = {2001.08266},
 primaryClass = {astro-ph.SR},
       adsurl = {https://ui.adsabs.harvard.edu/abs/2020A&A...638A.103C}
}

@ARTICLE{2020A&A...633A..99C,
       author = {{Cantat-Gaudin}, T. and {Anders}, F.},
        title = "{Clusters and mirages: cataloguing stellar aggregates in the Milky Way}",
      journal = {\aap},
         year = 2020,
        month = jan,
       volume = {633},
          eid = {A99},
        pages = {A99},
          doi = {10.1051/0004-6361/201936691},
archivePrefix = {arXiv},
       eprint = {1911.07075},
 primaryClass = {astro-ph.SR},
       adsurl = {https://ui.adsabs.harvard.edu/abs/2020A&A...633A..99C}
}

@PHDTHESIS{2019PhDT........68F,
       author = {{Fragkou}, Vasiliki},
        title = "{Galactic planetary nebulae with emphasis on those that are members of open star clusters.}",
       school = {University of Hong Kong},
         year = 2019,
        month = oct,
       adsurl = {https://ui.adsabs.harvard.edu/abs/2019PhDT........68F}
}

@ARTICLE{2019ARA&A..57..227K,
       author = {{Krumholz}, Mark R. and {McKee}, Christopher F. and {Bland-Hawthorn}, Joss},
        title = "{Star Clusters Across Cosmic Time}",
      journal = {\araa},
         year = 2019,
        month = aug,
       volume = {57},
        pages = {227-303},
          doi = {10.1146/annurev-astro-091918-104430},
archivePrefix = {arXiv},
       eprint = {1812.01615},
 primaryClass = {astro-ph.GA},
       adsurl = {https://ui.adsabs.harvard.edu/abs/2019ARA&A..57..227K}
}

@ARTICLE{2019NatAs...3..851F,
       author = {{Fragkou}, V. and {Parker}, Q.~A. and {Zijlstra}, A.~A. and {Crause}, L. and {Barker}, H.},
        title = "{A high-mass planetary nebula in a Galactic open cluster}",
      journal = {Nature Astronomy},
         year = 2019,
        month = jun,
       volume = {3},
        pages = {851-857},
          doi = {10.1038/s41550-019-0796-x},
archivePrefix = {arXiv},
       eprint = {1906.10556},
 primaryClass = {astro-ph.SR},
       adsurl = {https://ui.adsabs.harvard.edu/abs/2019NatAs...3..851F}
}

@ARTICLE{2019MNRAS.484.3078F,
       author = {{Fragkou}, V. and {Parker}, Q.~A. and {Zijlstra}, A. and {Shaw}, R. and {Lykou}, F.},
        title = "{The central star of planetary nebula PHR 1315 - 6555 and its host Galactic open cluster AL 1}",
      journal = {\mnras},
         year = 2019,
        month = apr,
       volume = {484},
       number = {3},
        pages = {3078-3092},
          doi = {10.1093/mnras/stz108},
archivePrefix = {arXiv},
       eprint = {1901.04174},
 primaryClass = {astro-ph.SR},
       adsurl = {https://ui.adsabs.harvard.edu/abs/2019MNRAS.484.3078F}
}

@ARTICLE{2019A&A...623A..80C,
       author = {{Carrera}, R. and {Bragaglia}, A. and {Cantat-Gaudin}, T. and {Vallenari}, A. and {Balaguer-N{\'u}{\~n}ez}, L. and {Bossini}, D. and {Casamiquela}, L. and {Jordi}, C. and {Sordo}, R. and {Soubiran}, C.},
        title = "{Open clusters in APOGEE and GALAH. Combining Gaia and ground-based spectroscopic surveys}",
      journal = {\aap},
         year = 2019,
        month = mar,
       volume = {623},
          eid = {A80},
        pages = {A80},
          doi = {10.1051/0004-6361/201834546},
archivePrefix = {arXiv},
       eprint = {1901.09302},
 primaryClass = {astro-ph.GA},
       adsurl = {https://ui.adsabs.harvard.edu/abs/2019A&A...623A..80C}
}

@ARTICLE{2019MNRAS.482.5138B,
       author = {{Baumgardt}, H. and {Hilker}, M. and {Sollima}, A. and {Bellini}, A.},
        title = "{Mean proper motions, space orbits, and velocity dispersion profiles of Galactic globular clusters derived from Gaia DR2 data}",
      journal = {\mnras},
         year = 2019,
        month = feb,
       volume = {482},
       number = {4},
        pages = {5138-5155},
          doi = {10.1093/mnras/sty2997},
archivePrefix = {arXiv},
       eprint = {1811.01507},
 primaryClass = {astro-ph.GA},
       adsurl = {https://ui.adsabs.harvard.edu/abs/2019MNRAS.482.5138B}
}

@ARTICLE{2018ApJ...866...21C,
       author = {{Cummings}, Jeffrey D. and {Kalirai}, Jason S. and {Tremblay}, P.-E. and {Ramirez-Ruiz}, Enrico and {Choi}, Jieun},
        title = "{The White Dwarf Initial-Final Mass Relation for Progenitor Stars from 0.85 to 7.5 M $_{{\ensuremath{\odot}}}$}",
      journal = {\apj},
         year = 2018,
        month = oct,
       volume = {866},
       number = {1},
          eid = {21},
        pages = {21},
          doi = {10.3847/1538-4357/aadfd6},
archivePrefix = {arXiv},
       eprint = {1809.01673},
 primaryClass = {astro-ph.SR},
       adsurl = {https://ui.adsabs.harvard.edu/abs/2018ApJ...866...21C}
}

@ARTICLE{2018A&A...616A...2L,
       author = {{Lindegren}, L. and {Hern{\'a}ndez}, J. and {Bombrun}, A. and {Klioner}, S. and {Bastian}, U. and {Ramos-Lerate}, M. and {de Torres}, A. and {Steidelm{\"u}ller}, H. and {Stephenson}, C. and {Hobbs}, D. and {Lammers}, U. and {Biermann}, M. and {Geyer}, R. and {Hilger}, T. and {Michalik}, D. and {Stampa}, U. and {McMillan}, P.~J. and {Casta{\~n}eda}, J. and {Clotet}, M. and {Comoretto}, G. and {Davidson}, M. and {Fabricius}, C. and {Gracia}, G. and {Hambly}, N.~C. and {Hutton}, A. and {Mora}, A. and {Portell}, J. and {van Leeuwen}, F. and {Abbas}, U. and {Abreu}, A. and {Altmann}, M. and {Andrei}, A. and {Anglada}, E. and {Balaguer-N{\'u}{\~n}ez}, L. and {Barache}, C. and {Becciani}, U. and {Bertone}, S. and {Bianchi}, L. and {Bouquillon}, S. and {Bourda}, G. and {Br{\"u}semeister}, T. and {Bucciarelli}, B. and {Busonero}, D. and {Buzzi}, R. and {Cancelliere}, R. and {Carlucci}, T. and {Charlot}, P. and {Cheek}, N. and {Crosta}, M. and {Crowley}, C. and {de Bruijne}, J. and {de Felice}, F. and {Drimmel}, R. and {Esquej}, P. and {Fienga}, A. and {Fraile}, E. and {Gai}, M. and {Garralda}, N. and {Gonz{\'a}lez-Vidal}, J.~J. and {Guerra}, R. and {Hauser}, M. and {Hofmann}, W. and {Holl}, B. and {Jordan}, S. and {Lattanzi}, M.~G. and {Lenhardt}, H. and {Liao}, S. and {Licata}, E. and {Lister}, T. and {L{\"o}ffler}, W. and {Marchant}, J. and {Martin-Fleitas}, J.-M. and {Messineo}, R. and {Mignard}, F. and {Morbidelli}, R. and {Poggio}, E. and {Riva}, A. and {Rowell}, N. and {Salguero}, E. and {Sarasso}, M. and {Sciacca}, E. and {Siddiqui}, H. and {Smart}, R.~L. and {Spagna}, A. and {Steele}, I. and {Taris}, F. and {Torra}, J. and {van Elteren}, A. and {van Reeven}, W. and {Vecchiato}, A.},
        title = "{Gaia Data Release 2. The astrometric solution}",
      journal = {\aap},
         year = 2018,
        month = aug,
       volume = {616},
          eid = {A2},
        pages = {A2},
          doi = {10.1051/0004-6361/201832727},
archivePrefix = {arXiv},
       eprint = {1804.09366},
 primaryClass = {astro-ph.IM},
       adsurl = {https://ui.adsabs.harvard.edu/abs/2018A&A...616A...2L}
}

@ARTICLE{2017ApJ...836...93J,
       author = {{Jacoby}, George H. and {De Marco}, Orsola and {Davies}, James and {Lotarevich}, I. and {Bond}, Howard E. and {Harrington}, J. Patrick and {Lanz}, Thierry},
        title = "{Masses of the Planetary Nebula Central Stars in the Galactic Globular Cluster System from HST Imaging and Spectroscopy}",
      journal = {\apj},
         year = 2017,
        month = feb,
       volume = {836},
       number = {1},
          eid = {93},
        pages = {93},
          doi = {10.3847/1538-4357/836/1/93},
archivePrefix = {arXiv},
       eprint = {1701.03516},
 primaryClass = {astro-ph.SR},
       adsurl = {https://ui.adsabs.harvard.edu/abs/2017ApJ...836...93J}
}

@INPROCEEDINGS{2016JPhCS.728c2008P,
       author = {{Parker}, Quentin A. and {Boji{\v{c}}i{\'c}}, Ivan S. and {Frew}, David J.},
        title = "{HASH: the Hong Kong/AAO/Strasbourg H{\ensuremath{\alpha}} planetary nebula database}",
    booktitle = {Journal of Physics Conference Series},
         year = 2016,
       series = {Journal of Physics Conference Series},
       volume = {728},
        month = jul,
    publisher = {IOP},
          eid = {032008},
        pages = {032008},
          doi = {10.1088/1742-6596/728/3/032008},
archivePrefix = {arXiv},
       eprint = {1603.07042},
 primaryClass = {astro-ph.SR},
       adsurl = {https://ui.adsabs.harvard.edu/abs/2016JPhCS.728c2008P}
}

@ARTICLE{2016A&A...588A..25M,
       author = {{Miller Bertolami}, Marcelo Miguel},
        title = "{New models for the evolution of post-asymptotic giant branch stars and central stars of planetary nebulae}",
      journal = {\aap},
         year = 2016,
        month = apr,
       volume = {588},
          eid = {A25},
        pages = {A25},
          doi = {10.1051/0004-6361/201526577},
archivePrefix = {arXiv},
       eprint = {1410.1679},
 primaryClass = {astro-ph.SR},
       adsurl = {https://ui.adsabs.harvard.edu/abs/2016A&A...588A..25M}
}

@ARTICLE{2016MNRAS.455.1459F,
       author = {{Frew}, David J. and {Parker}, Q.~A. and {Boji{\v{c}}i{\'c}}, I.~S.},
        title = "{The H{\ensuremath{\alpha}} surface brightness-radius relation: a robust statistical distance indicator for planetary nebulae}",
      journal = {\mnras},
         year = 2016,
        month = jan,
       volume = {455},
       number = {2},
        pages = {1459-1488},
          doi = {10.1093/mnras/stv1516},
archivePrefix = {arXiv},
       eprint = {1504.01534},
 primaryClass = {astro-ph.SR},
       adsurl = {https://ui.adsabs.harvard.edu/abs/2016MNRAS.455.1459F}
}

@ARTICLE{2015A&A...573A..42L,
       author = {{Luridiana}, V. and {Morisset}, C. and {Shaw}, R.~A.},
        title = "{PyNeb: a new tool for analyzing emission lines. I. Code description and validation of results}",
      journal = {\aap},
         year = 2015,
        month = jan,
       volume = {573},
          eid = {A42},
        pages = {A42},
          doi = {10.1051/0004-6361/201323152},
archivePrefix = {arXiv},
       eprint = {1410.6662},
 primaryClass = {astro-ph.IM},
       adsurl = {https://ui.adsabs.harvard.edu/abs/2015A&A...573A..42L}
}

@ARTICLE{2013A&A...558A..53K,
       author = {{Kharchenko}, N.~V. and {Piskunov}, A.~E. and {Schilbach}, E. and {R{\"o}ser}, S. and {Scholz}, R.-D.},
        title = "{Global survey of star clusters in the Milky Way. II. The catalogue of basic parameters}",
      journal = {\aap},
         year = 2013,
        month = oct,
       volume = {558},
          eid = {A53},
        pages = {A53},
          doi = {10.1051/0004-6361/201322302},
archivePrefix = {arXiv},
       eprint = {1308.5822},
 primaryClass = {astro-ph.GA},
       adsurl = {https://ui.adsabs.harvard.edu/abs/2013A&A...558A..53K}
}

@ARTICLE{2013ApJ...775..134V,
       author = {{VandenBerg}, Don A. and {Brogaard}, K. and {Leaman}, R. and {Casagrande}, L.},
        title = "{The Ages of 55 Globular Clusters as Determined Using an Improved \textbackslashDelta V\^HB\_TO Method along with Color-Magnitude Diagram Constraints, and Their Implications for Broader Issues}",
      journal = {\apj},
         year = 2013,
        month = oct,
       volume = {775},
       number = {2},
          eid = {134},
        pages = {134},
          doi = {10.1088/0004-637X/775/2/134},
archivePrefix = {arXiv},
       eprint = {1308.2257},
 primaryClass = {astro-ph.GA},
       adsurl = {https://ui.adsabs.harvard.edu/abs/2013ApJ...775..134V}
}

@ARTICLE{2013MNRAS.431....2F,
       author = {{Frew}, David J. and {Boji{\v{c}}i{\'c}}, Ivan S. and {Parker}, Q.~A.},
        title = "{A catalogue of integrated H{\ensuremath{\alpha}} fluxes for 1258 Galactic planetary nebulae}",
      journal = {\mnras},
         year = 2013,
        month = may,
       volume = {431},
       number = {1},
        pages = {2-26},
          doi = {10.1093/mnras/sts393},
archivePrefix = {arXiv},
       eprint = {1211.2505},
 primaryClass = {astro-ph.SR},
       adsurl = {https://ui.adsabs.harvard.edu/abs/2013MNRAS.431....2F}
}

@ARTICLE{2013ApJ...765L..43I,
       author = {{Ibeling}, Duligur and {Heger}, Alexander},
        title = "{The Metallicity Dependence of the Minimum Mass for Core-collapse Supernovae}",
      journal = {\apjl},
         year = 2013,
        month = mar,
       volume = {765},
       number = {2},
          eid = {L43},
        pages = {L43},
          doi = {10.1088/2041-8205/765/2/L43},
archivePrefix = {arXiv},
       eprint = {1301.5783},
 primaryClass = {astro-ph.SR},
       adsurl = {https://ui.adsabs.harvard.edu/abs/2013ApJ...765L..43I}
}

@ARTICLE{2012MNRAS.427..127B,
       author = {{Bressan}, Alessandro and {Marigo}, Paola and {Girardi}, L{\'e}o. and {Salasnich}, Bernardo and {Dal Cero}, Claudia and {Rubele}, Stefano and {Nanni}, Ambra},
        title = "{PARSEC: stellar tracks and isochrones with the PAdova and TRieste Stellar Evolution Code}",
      journal = {\mnras},
         year = 2012,
        month = nov,
       volume = {427},
       number = {1},
        pages = {127-145},
          doi = {10.1111/j.1365-2966.2012.21948.x},
archivePrefix = {arXiv},
       eprint = {1208.4498},
 primaryClass = {astro-ph.SR},
       adsurl = {https://ui.adsabs.harvard.edu/abs/2012MNRAS.427..127B}
}

@ARTICLE{2011MNRAS.413.1835P,
       author = {{Parker}, Quentin A. and {Frew}, David J. and {Miszalski}, Brent and {Kovacevic}, Anna V. and {Frinchaboy}, Peter M. and {Dobbie}, Paul D. and {K{\"o}ppen}, Joachim},
        title = "{PHR 1315-6555: a bipolar planetary nebula in the compact Hyades-age open cluster ESO 96-SC04}",
      journal = {\mnras},
         year = 2011,
        month = may,
       volume = {413},
       number = {3},
        pages = {1835-1844},
          doi = {10.1111/j.1365-2966.2011.18259.x},
archivePrefix = {arXiv},
       eprint = {1101.3814},
 primaryClass = {astro-ph.GA},
       adsurl = {https://ui.adsabs.harvard.edu/abs/2011MNRAS.413.1835P}
}

@ARTICLE{2011ApJ...730...67B,
       author = {{Brown}, Justin M. and {Kilic}, Mukremin and {Brown}, Warren R. and {Kenyon}, Scott J.},
        title = "{The Binary Fraction of Low-mass White Dwarfs}",
      journal = {\apj},
         year = 2011,
        month = apr,
       volume = {730},
       number = {2},
          eid = {67},
        pages = {67},
          doi = {10.1088/0004-637X/730/2/67},
archivePrefix = {arXiv},
       eprint = {1101.5169},
 primaryClass = {astro-ph.GA},
       adsurl = {https://ui.adsabs.harvard.edu/abs/2011ApJ...730...67B}
}

@ARTICLE{2011CQGra..28h5016W,
       author = {{White}, Darren J. and {Daw}, E.~J. and {Dhillon}, V.~S.},
        title = "{A list of galaxies for gravitational wave searches}",
      journal = {Classical and Quantum Gravity},
         year = 2011,
        month = apr,
       volume = {28},
       number = {8},
          eid = {085016},
        pages = {085016},
          doi = {10.1088/0264-9381/28/8/085016},
archivePrefix = {arXiv},
       eprint = {1103.0695},
 primaryClass = {astro-ph.CO},
       adsurl = {https://ui.adsabs.harvard.edu/abs/2011CQGra..28h5016W}
}

@ARTICLE{2010AJ....139.2440R,
       author = {{Roeser}, S. and {Demleitner}, M. and {Schilbach}, E.},
        title = "{The PPMXL Catalog of Positions and Proper Motions on the ICRS. Combining USNO-B1.0 and the Two Micron All Sky Survey (2MASS)}",
      journal = {\aj},
         year = 2010,
        month = jun,
       volume = {139},
       number = {6},
        pages = {2440-2447},
          doi = {10.1088/0004-6256/139/6/2440},
archivePrefix = {arXiv},
       eprint = {1003.5852},
 primaryClass = {astro-ph.GA},
       adsurl = {https://ui.adsabs.harvard.edu/abs/2010AJ....139.2440R}
}

@ARTICLE{2009PASP..121...76C,
       author = {{Carey}, S.~J. and {Noriega-Crespo}, A. and {Mizuno}, D.~R. and {Shenoy}, S. and {Paladini}, R. and {Kraemer}, K.~E. and {Price}, S.~D. and {Flagey}, N. and {Ryan}, E. and {Ingalls}, J.~G. and {Kuchar}, T.~A. and {Pinheiro Gon{\c{c}}alves}, Daniela and {Indebetouw}, R. and {Billot}, N. and {Marleau}, F.~R. and {Padgett}, D.~L. and {Rebull}, L.~M. and {Bressert}, E. and {Ali}, Babar and {Molinari}, S. and {Martin}, P.~G. and {Berriman}, G.~B. and {Boulanger}, F. and {Latter}, W.~B. and {Miville-Deschenes}, M.~A. and {Shipman}, R. and {Testi}, L.},
        title = "{MIPSGAL: A Survey of the Inner Galactic Plane at 24 and 70 {\ensuremath{\mu}}m}",
      journal = {\pasp},
         year = 2009,
        month = jan,
       volume = {121},
       number = {875},
        pages = {76},
          doi = {10.1086/596581},
       adsurl = {https://ui.adsabs.harvard.edu/abs/2009PASP..121...76C}
}

@ARTICLE{2007CSE.....9...90H,
       author = {{Hunter}, John D.},
        title = "{Matplotlib: A 2D Graphics Environment}",
      journal = {Computing in Science and Engineering},
         year = 2007,
        month = jan,
       volume = {9},
       number = {3},
        pages = {90-95},
          doi = {10.1109/MCSE.2007.55},
       adsurl = {https://ui.adsabs.harvard.edu/abs/2007CSE.....9...90H}
}

@ARTICLE{2006AJ....131.1163S,
       author = {{Skrutskie}, M.~F. and {Cutri}, R.~M. and {Stiening}, R. and {Weinberg}, M.~D. and {Schneider}, S. and {Carpenter}, J.~M. and {Beichman}, C. and {Capps}, R. and {Chester}, T. and {Elias}, J. and {Huchra}, J. and {Liebert}, J. and {Lonsdale}, C. and {Monet}, D.~G. and {Price}, S. and {Seitzer}, P. and {Jarrett}, T. and {Kirkpatrick}, J.~D. and {Gizis}, J.~E. and {Howard}, E. and {Evans}, T. and {Fowler}, J. and {Fullmer}, L. and {Hurt}, R. and {Light}, R. and {Kopan}, E.~L. and {Marsh}, K.~A. and {McCallon}, H.~L. and {Tam}, R. and {Van Dyk}, S. and {Wheelock}, S.},
        title = "{The Two Micron All Sky Survey (2MASS)}",
      journal = {\aj},
         year = 2006,
        month = feb,
       volume = {131},
       number = {2},
        pages = {1163-1183},
          doi = {10.1086/498708},
       adsurl = {https://ui.adsabs.harvard.edu/abs/2006AJ....131.1163S}
}

@INPROCEEDINGS{2005ASPC..347...29T,
       author = {{Taylor}, M.~B.},
        title = "{TOPCAT \& STIL: Starlink Table/VOTable Processing Software}",
    booktitle = {Astronomical Data Analysis Software and Systems XIV},
         year = 2005,
       editor = {{Shopbell}, P. and {Britton}, M. and {Ebert}, R.},
       series = {Astronomical Society of the Pacific Conference Series},
       volume = {347},
        month = dec,
        pages = {29},
       adsurl = {https://ui.adsabs.harvard.edu/abs/2005ASPC..347...29T}
}

@BOOK{2000oepn.book.....K,
       author = {{Kwok}, Sun},
        title = "{The Origin and Evolution of Planetary Nebulae}",
         year = 2000,
       adsurl = {https://ui.adsabs.harvard.edu/abs/2000oepn.book.....K}
}

@INPROCEEDINGS{2000ASPC..198..517M,
       author = {{Mathieu}, R.~D.},
        title = "{The WIYN Open Cluster Study}",
    booktitle = {Stellar Clusters and Associations: Convection, Rotation, and Dynamos},
         year = 2000,
       editor = {{Pallavicini}, R. and {Micela}, G. and {Sciortino}, S.},
       series = {Astronomical Society of the Pacific Conference Series},
       volume = {198},
        month = jan,
        pages = {517},
       adsurl = {https://ui.adsabs.harvard.edu/abs/2000ASPC..198..517M}
}

@ARTICLE{1997AJ....114.2611J,
       author = {{Jacoby}, George H. and {Morse}, Jon A. and {Fullton}, L. Kellar and {Kwitter}, K.~B. and {Henry}, R.~B.~C.},
        title = "{Planetary Nebulae in the Globular Cluster PAL 6 and NGC 6441}",
      journal = {\aj},
         year = 1997,
        month = dec,
       volume = {114},
        pages = {2611},
          doi = {10.1086/118671},
       adsurl = {https://ui.adsabs.harvard.edu/abs/1997AJ....114.2611J}
}

@ARTICLE{1996AJ....112.1487H,
       author = {{Harris}, William E.},
        title = "{A Catalog of Parameters for Globular Clusters in the Milky Way}",
      journal = {\aj},
         year = 1996,
        month = oct,
       volume = {112},
        pages = {1487},
          doi = {10.1086/118116},
       adsurl = {https://ui.adsabs.harvard.edu/abs/1996AJ....112.1487H}
}

@ARTICLE{1996ApJ...460..914V,
       author = {{Vacca}, William D. and {Garmany}, Catharine D. and {Shull}, J. Michael},
        title = "{The Lyman-Continuum Fluxes and Stellar Parameters of O and Early B-Type Stars}",
      journal = {\apj},
         year = 1996,
        month = apr,
       volume = {460},
        pages = {914},
          doi = {10.1086/177020},
       adsurl = {https://ui.adsabs.harvard.edu/abs/1996ApJ...460..914V}
}

@ARTICLE{1989ApJ...345..245C,
       author = {{Cardelli}, Jason A. and {Clayton}, Geoffrey C. and {Mathis}, John S.},
        title = "{The Relationship between Infrared, Optical, and Ultraviolet Extinction}",
      journal = {\apj},
         year = 1989,
        month = oct,
       volume = {345},
        pages = {245},
          doi = {10.1086/167900},
       adsurl = {https://ui.adsabs.harvard.edu/abs/1989ApJ...345..245C}
}

@ARTICLE{1989ApJ...338..862G,
       author = {{Gillett}, F.~C. and {Jacoby}, G.~H. and {Joyce}, R.~R. and {Cohen}, J.~G. and {Neugebauer}, G. and {Soifer}, B.~T. and {Nakajima}, T. and {Matthews}, K.},
        title = "{The Optical/Infrared Counterpart(s) of IRAS 18333-2357}",
      journal = {\apj},
         year = 1989,
        month = mar,
       volume = {338},
        pages = {862},
          doi = {10.1086/167241},
       adsurl = {https://ui.adsabs.harvard.edu/abs/1989ApJ...338..862G}
}

@INPROCEEDINGS{1983IAUS..103..233P,
       author = {{Peimbert}, M. and {Torres-Peimbert}, S.},
        title = "{Type I planetary nebulae.}",
    booktitle = {Planetary Nebulae},
         year = 1983,
       editor = {{Aller}, L.~H.},
       series = {IAU Symposium},
       volume = {103},
        month = jan,
        pages = {233-242},
       adsurl = {https://ui.adsabs.harvard.edu/abs/1983IAUS..103..233P}
}

@ARTICLE{1982ApJ...258..280K,
       author = {{Kwok}, S.},
        title = "{From red giants to planetary nebulae.}",
      journal = {\apj},
         year = 1982,
        month = jul,
       volume = {258},
        pages = {280-288},
          doi = {10.1086/160078},
       adsurl = {https://ui.adsabs.harvard.edu/abs/1982ApJ...258..280K}
}

@BOOK{1976asqu.book.....A,
       author = {{Allen}, C.~W.},
        title = "{Astrophysical Quantities}",
         year = 1976,
       adsurl = {https://ui.adsabs.harvard.edu/abs/1976asqu.book.....A}
}

@ARTICLE{1931ZA......2....1Z,
       author = {{Zanstra}, H.},
        title = "{Untersuchungen {\"u}ber planetarische Nebel. Erster Teil: Der Leuchtproze\&beta planetarischer Nebel und die Temperatur der Zentralsterne. Mit 4 Abbildungen.}",
      journal = {\zap},
         year = 1931,
        month = jan,
       volume = {2},
        pages = {1},
       adsurl = {https://ui.adsabs.harvard.edu/abs/1931ZA......2....1Z}
}

@ARTICLE{1928PASP...40..342P,
       author = {{Pease}, F.~G.},
        title = "{A Planetary Nebula in the Globular Cluster Messier 15}",
      journal = {\pasp},
         year = 1928,
        month = oct,
       volume = {40},
       number = {237},
        pages = {342-342},
          doi = {10.1086/123857},
       adsurl = {https://ui.adsabs.harvard.edu/abs/1928PASP...40..342P}
}

@ARTICLE{2024A&A...687A.214G,
       author = {{Garro}, E.~R. and {Minniti}, D. and {Fern{\'a}ndez-Trincado}, J.~G.},
        title = "{Over 200 globular clusters in the Milky Way and still none with super-Solar metallicity}",
      journal = {\aap},
         year = 2024,
        month = jul,
       volume = {687},
          eid = {A214},
        pages = {A214},
          doi = {10.1051/0004-6361/202347389},
archivePrefix = {arXiv},
       eprint = {2405.05055},
 primaryClass = {astro-ph.GA},
       adsurl = {https://ui.adsabs.harvard.edu/abs/2024A&A...687A.214G}
}

@ARTICLE{2024A&A...687A.201B,
       author = {{Bica}, E. and {Ortolani}, S. and {Barbuy}, B. and {Oliveira}, R.~A.~P.},
        title = "{A census of new globular clusters in the Galactic bulge}",
      journal = {\aap},
         year = 2024,
        month = jul,
       volume = {687},
          eid = {A201},
        pages = {A201},
          doi = {10.1051/0004-6361/202346377},
archivePrefix = {arXiv},
       eprint = {2405.03068},
 primaryClass = {astro-ph.GA},
       adsurl = {https://ui.adsabs.harvard.edu/abs/2024A&A...687A.201B}
}

@ARTICLE{2024AJ....168..160B,
       author = {{Bond}, Howard E. and {Bellini}, Andrea and {Sahu}, Kailash C.},
        title = "{Testing Cluster Membership of Planetary Nebulae with High-precision Proper Motions. I. HST Observations of JaFu 1 Near the Globular Cluster Palomar 6}",
      journal = {\aj},
         year = 2024,
        month = oct,
       volume = {168},
       number = {4},
          eid = {160},
        pages = {160},
          doi = {10.3847/1538-3881/ad67d4},
archivePrefix = {arXiv},
       eprint = {2407.18135},
 primaryClass = {astro-ph.SR},
       adsurl = {https://ui.adsabs.harvard.edu/abs/2024AJ....168..160B}
}

@ARTICLE{2013ApJ...769...10J,
       author = {{Jacoby}, George H. and {Ciardullo}, Robin and {De Marco}, Orsola and {Lee}, Myung Gyoon and {Herrmann}, Kimberly A. and {Hwang}, Ho Seong and {Kaplan}, Evan and {Davies}, James E.},
        title = "{A Survey for Planetary Nebulae in M31 Globular Clusters}",
      journal = {\apj},
         year = 2013,
        month = may,
       volume = {769},
       number = {1},
          eid = {10},
        pages = {10},
          doi = {10.1088/0004-637X/769/1/10},
archivePrefix = {arXiv},
       eprint = {1303.3867},
 primaryClass = {astro-ph.CO},
       adsurl = {https://ui.adsabs.harvard.edu/abs/2013ApJ...769...10J}
}

@ARTICLE{2002ApJ...575L..59M,
       author = {{Minniti}, Dante and {Rejkuba}, Marina},
        title = "{Extragalactic Globular Cluster Planetary Nebulae: Discovery of a Planetary Nebula in the NGC 5128 Globular Cluster G169 Using the Magellan I Baade Telescope}",
      journal = {\apjl},
         year = 2002,
        month = aug,
       volume = {575},
       number = {2},
        pages = {L59-L62},
          doi = {10.1086/342827},
       adsurl = {https://ui.adsabs.harvard.edu/abs/2002ApJ...575L..59M}
}

@ARTICLE{2014A&A...561A.119M,
       author = {{Moni Bidin}, C. and {Majaess}, D. and {Bonatto}, C. and {Mauro}, F. and {Turner}, D. and {Geisler}, D. and {Chen{\'e}}, A.-N. and {Gormaz-Matamala}, A.~C. and {Borissova}, J. and {Kurtev}, R.~G. and {Minniti}, D. and {Carraro}, G. and {Gieren}, W.},
        title = "{Investigating potential planetary nebula/cluster pairs}",
      journal = {\aap},
         year = 2014,
        month = jan,
       volume = {561},
          eid = {A119},
        pages = {A119},
          doi = {10.1051/0004-6361/201220802},
archivePrefix = {arXiv},
       eprint = {1311.0760},
 primaryClass = {astro-ph.GA},
       adsurl = {https://ui.adsabs.harvard.edu/abs/2014A&A...561A.119M}
}

@ARTICLE{2019ApJ...884L..15M,
       author = {{Minniti}, Dante and {Dias}, Bruno and {G{\'o}mez}, Mat{\'\i}as and {Palma}, Tali and {Pullen}, Joyce B.},
        title = "{New Candidate Planetary Nebulae in Galactic Globular Clusters from the VVV Survey}",
      journal = {\apjl},
         year = 2019,
        month = oct,
       volume = {884},
       number = {1},
          eid = {L15},
        pages = {L15},
          doi = {10.3847/2041-8213/ab4424},
archivePrefix = {arXiv},
       eprint = {1909.09109},
 primaryClass = {astro-ph.SR},
       adsurl = {https://ui.adsabs.harvard.edu/abs/2019ApJ...884L..15M}
}

@ARTICLE{2008A&A...477L..17L,
       author = {{Larsen}, S.~S.},
        title = "{A peculiar planetary nebula candidate in a globular cluster in the Fornax dwarf spheroidal galaxy}",
      journal = {\aap},
         year = 2008,
        month = jan,
       volume = {477},
       number = {2},
        pages = {L17-L20},
          doi = {10.1051/0004-6361:20078950},
archivePrefix = {arXiv},
       eprint = {0711.2249},
 primaryClass = {astro-ph},
       adsurl = {https://ui.adsabs.harvard.edu/abs/2008A&A...477L..17L}
}

@ARTICLE{2025PASP..137k4202B,
       author = {{Bond}, Howard E. and {Bastian}, Nate and {Bellini}, Andrea and {Kamann}, Sebastian and {Libralato}, Mattia and {Niederhofer}, Florian and {Roth}, Martin M. and {Soemitro}, Azlizan A.},
        title = "{Serendipitous Discovery of a Faint Planetary Nebula in the Massive Young LMC Cluster NGC 1866}",
      journal = {\pasp},
         year = 2025,
        month = nov,
       volume = {137},
       number = {11},
          eid = {114202},
        pages = {114202},
          doi = {10.1088/1538-3873/ae1664},
archivePrefix = {arXiv},
       eprint = {2510.16630},
 primaryClass = {astro-ph.SR},
       adsurl = {https://ui.adsabs.harvard.edu/abs/2025PASP..137k4202B}
}

@ARTICLE{2021A&A...650L..11M,
       author = {{Minniti}, Dante and {Fern{\'a}ndez-Trincado}, Jos{\'e} G. and {G{\'o}mez}, Mat{\'\i}as and {Smith}, Leigh C. and {Lucas}, Philip W. and {Contreras Ramos}, Rodrigo},
        title = "{Discovery of a new nearby globular cluster with extreme kinematics located in the extension of a halo stream}",
      journal = {\aap},
         year = 2021,
        month = jun,
       volume = {650},
          eid = {L11},
        pages = {L11},
          doi = {10.1051/0004-6361/202141129},
archivePrefix = {arXiv},
       eprint = {2106.01383},
 primaryClass = {astro-ph.GA},
       adsurl = {https://ui.adsabs.harvard.edu/abs/2021A&A...650L..11M}
}

@ARTICLE{2003ARA&A..41...57L,
       author = {{Lada}, Charles J. and {Lada}, Elizabeth A.},
        title = "{Embedded Clusters in Molecular Clouds}",
      journal = {\araa},
         year = 2003,
        month = jan,
       volume = {41},
        pages = {57-115},
          doi = {10.1146/annurev.astro.41.011802.094844},
archivePrefix = {arXiv},
       eprint = {astro-ph/0301540},
 primaryClass = {astro-ph},
       adsurl = {https://ui.adsabs.harvard.edu/abs/2003ARA&A..41...57L}
}

@ARTICLE{2019ApJ...884..115D,
       author = {{Davis}, Brian D. and {Bond}, Howard E. and {Ciardullo}, Robin and {Jacoby}, George H.},
        title = "{Hubble Space Telescope Spectroscopy of a Planetary Nebula in an M31 Open Cluster: Hot-bottom Burning at 3.4 M $_{☉}$}",
      journal = {\apj},
         year = 2019,
        month = oct,
       volume = {884},
       number = {2},
          eid = {115},
        pages = {115},
          doi = {10.3847/1538-4357/ab44d4},
archivePrefix = {arXiv},
       eprint = {1909.08007},
 primaryClass = {astro-ph.SR},
       adsurl = {https://ui.adsabs.harvard.edu/abs/2019ApJ...884..115D}
}

@ARTICLE{1998A&AS..132...13D,
       author = {{Durand}, S. and {Acker}, A. and {Zijlstra}, A.},
        title = "{The kinematics of 867 galactic planetary nebulae}",
      journal = {\aaps},
         year = 1998,
        month = oct,
       volume = {132},
        pages = {13-20},
          doi = {10.1051/aas:1998356},
       adsurl = {https://ui.adsabs.harvard.edu/abs/1998A&AS..132...13D}
}

@ARTICLE{2020AJ....159..276B,
       author = {{Bond}, Howard E. and {Bellini}, Andrea and {Sahu}, Kailash C.},
        title = "{Proper-motion Membership Tests for Four Planetary Nebulae in Galactic Globular Clusters}",
      journal = {\aj},
         year = 2020,
        month = jun,
       volume = {159},
       number = {6},
          eid = {276},
        pages = {276},
          doi = {10.3847/1538-3881/ab8f9e},
archivePrefix = {arXiv},
       eprint = {2002.11653},
 primaryClass = {astro-ph.SR},
       adsurl = {https://ui.adsabs.harvard.edu/abs/2020AJ....159..276B}
}

@ARTICLE{2009PASP..121..316D,
       author = {{De Marco}, Orsola},
        title = "{The Origin and Shaping of Planetary Nebulae: Putting the Binary Hypothesis to the Test}",
      journal = {\pasp},
         year = 2009,
        month = apr,
       volume = {121},
       number = {878},
        pages = {316},
          doi = {10.1086/597765},
archivePrefix = {arXiv},
       eprint = {0902.1137},
 primaryClass = {astro-ph.GA},
       adsurl = {https://ui.adsabs.harvard.edu/abs/2009PASP..121..316D}
}

@ARTICLE{2000A&AS..143...23O,
       author = {{Ochsenbein}, F. and {Bauer}, P. and {Marcout}, J.},
        title = "{The VizieR database of astronomical catalogues}",
      journal = {\aaps},
         year = 2000,
        month = apr,
       volume = {143},
        pages = {23-32},
          doi = {10.1051/aas:2000169},
archivePrefix = {arXiv},
       eprint = {astro-ph/0002122},
 primaryClass = {astro-ph},
       adsurl = {https://ui.adsabs.harvard.edu/abs/2000A&AS..143...23O}
}
\bibliographystyle{aasjournalv7}

\end{document}